\documentclass[10pt]{article}\usepackage{amsmath,amssymb,multirow,jheppub,graphicx}
\usepackage{centernot}
\usepackage{mciteplus}
\usepackage[utf8]{inputenc}
\usepackage[makeroom]{cancel}
\usepackage{tikz}
\usepackage[caption=false]{subfig}

\usepackage{color}
\allowdisplaybreaks[1]
\usepackage{tikz}
 \tikzset{node distance=2cm, auto}

\renewcommand{\Im}{\text{Im }}

\def\Im{\text{Im}}

\def\tr{\text{tr}}

\def\hat{\widehat}
\def\bar{\overline}

\def\I{{\cal I}}

\def\coeff#1#2{{\textstyle {\frac {#1}{#2}}}}

\def\half{\coeff 12}

\usepackage{verbatim}   
\usepackage[all]{xy}
\usepackage{bm}  
\def\Dslash{{\rlap{\raise 1pt \hbox{$\>/$}}D}}
\def\Pslash{{\rlap{\raise  1pt \hbox{$\>/$}}\,\partial}}

\newcommand{\be}{\begin{equation}}      
\newcommand{\ee}{\end{equation}}      
\newcommand{\bea}{\begin{eqnarray}}      
\newcommand{\eea}{\end{eqnarray}}

\title{
Cheshire Cat resurgence, Self-resurgence and Quasi-Exact Solvable Systems}

\author[1,2]{Can Koz\c{c}az}\author[3]{Tin Sulejmanpasic}\author[4]{Yuya Tanizaki}   \author[3]{Mithat  \"Unsal}
\affiliation[1]{Center of Mathematical Sciences and Applications, Harvard University, Cambridge, MA 02138, USA}
\affiliation[2]{Jefferson Physical Laboratory, Harvard University, Cambridge, MA 02138, USA}
\affiliation[3]{Department of Physics, North Carolina State University, Raleigh, NC, 27695}
\affiliation[4]{RIKEN BNL Research Center, Brookhaven National Laboratory, Upton, NY 11973 US}
\emailAdd{kozcaz@cmsa.fas.harvard.edu,
tin.sulejmanpasic@gmail.com, tanizaki@bnl.gov, unsal.mithat@gmail.com}

\abstract{We explore a one parameter $\zeta$-deformation of the quantum-mechanical Sine-Gordon and Double-Well potentials which we call the Double Sine-Gordon (DSG) and the Tilted Double Well (TDW), respectively. In these systems, for positive integer values of $\zeta$, the lowest $\zeta$ states turn out to be exactly solvable for DSG -- a feature known as Quasi-Exact-Solvability (QES) -- and solvable to all orders in perturbation theory for TDW. For DSG such states do not show any instanton-like dependence on the coupling constant, although the action has real saddles. On the other hand, although it has no real saddles, the TDW admits all-orders perturbative states that are not normalizable, and hence, requires a non-perturbative energy shift. Both of these puzzles are solved by including complex saddles. We show that the convergence is dictated by the quantization of the hidden topological angle. Further, we argue that the QES systems can be linked to the exact cancellation of real and complex non-perturbative saddles to all orders in the semi-classical expansion. We also show that the entire resurgence structure remains encoded in the analytic properties of the $\zeta$-deformation, even though exactly at integer values of $\zeta$ the mechanism of resurgence is obscured by the lack of ambiguity in both the Borel sum of the perturbation theory as well as the non-perturbative contributions. In this way, all of the characteristics of resurgence remains even when its role seems to vanish, much like the lingering grin of the Cheshire Cat. We also show that the perturbative series is Self-resurgent -- a feature by which there is a one-to-one relation between the early terms of the perturbative expansion and the late terms of \emph{the same expansion} -- which is intimately connected with the Dunne-\"{U}nsal relation. We explicitly  verify that this is indeed the case.    }

\begin{document}
\maketitle

\newpage
\section{Introduction and Results}

For many quantum systems, perturbation theory is widely employed successfully to obtain approximate results. Following the immense success of the perturbative treatment of Quantum Electrodynamics which resulted in a Nobel Prize shared by Tomonaga, Schwinger and Feynman in 1965, the perturbation theory and Feynman diagrams became firmly associated with \emph{the Quantum Field Theory}. Indeed, concepts such as renormalization group, Bjorken scaling, running coupling and asymptotic freedom are just some of the concepts intimately tied to the utility and indispensability of the perturbation theory. 

 Nevertheless, already as early as 1952, Dyson gave a physical argument that analytic continuation of the electric charge $e^2\rightarrow -e^2$ would cause an instability, effectively indicating that in Quantum Electrodynamics --- the simplest and most accurately verified quantum field theory manifested in nature--- the radius of convergence of the perturbation theory is zero~\cite{Dyson:1952tj}. Since then it has become clear that this is a generic feature of both quantum mechanical as well as field theoretical systems,  with a typical divergence rate being factorial. It is for this reason that perturbation theory fails to define a quantum field theory, or even, indeed, quantum mechanics. 
 
Perhaps unsurprisingly while the successes of the perturbation theory are commonly praised and a matter of textbook knowledge, its apparent deficits deeply rooted in its structure are more often than not either overlooked or tacitly ignored. While this point of view is sometimes necessary and sometimes useful, it turns a blind eye to the beautiful and intricate structure hidden in the perturbative expansion and its stubborn insistence on diverging. The question then what, if any, is the meaning of such series.

\par{A celebrated way to make sense of factorially diverging series is to tame them by a special transformation--- the Borel transform, which renders the series convergent. The Laplace transform of this sum gives rise to another function--- the Borel sum--- which has the same asymptotic expansion as the original series but assigns a non-divergent value to it. If this can be done in a  unique way, the series is said to be \emph{Borel summable}, as there is a sense in which a divergent series is assigned a concrete value.
Still the Borel transform often has singularities in generic cases which may render the Borel sum ambiguous. The goal of resurgence theory is to describe the global nature of the solution by analyzing these singularities and ambiguities that they may cause ~\cite{Ecalle-book, Voros1983, candelpergher1993approche, delabaere1999resurgent, kawai2005algebraic}. 
For instance, if those singularities lie on the positive real axis, we might have to avoid these poles by going around them in the complex plane. 
Different deformations introduce imaginary ambiguities for physical observables, e.g. for energy levels. At first, we might be tempted to abandon this prescription due to this kind of pathological results once and for all. Nevertheless remarkably and perhaps surprisingly the pathology of the perturbation theory turns out to be inextricably linked to the non-perturbative physics~\cite{Bogomolny:1980ur,ZinnJustin:1981dx, Dunne:2014bca,Dunne:2013ada, Berry657,delabaere2002,Howls, Cherman:2014ofa, Behtash:2015loa, Misumi:2015dua}. In other words, the ambiguity caused by the divergence and non-Borel summability of the small coupling expansion serves as a placeholder, much like a pattern of a jigsaw puzzle, stitching perturbative and non-perturbative contributions in such a way to eliminate all ambiguities. The study, analysis and understanding of such phenomena are known under the name of \emph{resurgence theory}.}
\par{The resurgence theory developed by \'{E}calle~\cite{Ecalle-book} (in the context of non-linear differential equations)  is proficient enough to encode the subtle information around different saddles by replacing the conventional perturbation series with transseries. The transseries do not only consist of a power series in the coupling constant but also include non-analytic terms relevant to instanton contributions and the integration over their quasi-moduli. 
The power of the resurgence theory lies in the possibility that it may provide a consistent manner to take into account the presence of all saddle points under certain physical requirements. More concretely, it connects the perturbative fluctuations around different saddles via intricate relations with each other and respects the monodromy properties of the underlying quantum system. There has been an ever growing set of physical systems where resurgence theory resolves some puzzles and reveals surprises related to semi-classical analysis, such as the semi-classical interpretation and the role of renormalon-like singularities \cite{Argyres:2012ka,Argyres:2012vv,Dunne:2012zk,Dunne:2012ae,Cherman:2013yfa,Anber:2014sda}, the relation between perturbation theories among different saddles \cite{Dunne:2014bca,Basar:2015xna,Escobar-Ruiz:2015rfa,Escobar-Ruiz:2015nsa,DU2016},  stabilization of center symmetry in super Yang-Mills theory \cite{Poppitz:2012sw,Poppitz:2012nz}, Borel summability of $\mathcal N=2$ super Yang-Mills \cite{Argyres:2012ka,Argyres:2012vv,Honda:2016mvg}, the meaning and limits of the Bogomolny--Zinn-Justin prescription \cite{Behtash:2015kna,Behtash:2015kva}, the vanishing gluon condensate in SUSY gauge theories \cite{Behtash:2015kna}, the role of multi-instantons in $\mathcal N=1$ \cite{Behtash:2015kna,Behtash:2015zha} and $\mathcal N=2$ \cite{Behtash:2015kva} quantum mechanics, role of ``instantons''  and complex solutions in the Gross-Witten theory \cite{Buividovich:2015oju}, as well as an abundance of work ranging from quantum mechanics to general quantum field theory to string theory  ~\cite{Marino:2007te, Marino:2012zq, Aniceto:2011nu,Schiappa:2013opa,Santamaria:2013rua,Couso-Santamaria:2014iia,Vonk:2015sia,Couso-Santamaria:2016vcc,Aniceto:2013fka, Dunne:2013ada, Hatsuda:2013oxa, Cherman:2014ofa, Basar:2013eka, Misumi:2014jua, Misumi:2015dua, Misumi:2014bsa, Dorigoni:2014hea,Fujimori:2016ljw,Gukov:2016njj,Fitzpatrick:2016mjq,Fitzpatrick:2016ive, Demulder:2016mja}.   }
\par{In this work, we aim to solve yet another puzzle related to systems for which a part of the spectrum can be solved at isolated points in the parameter space. Such special systems are dubbed \emph{Quasi-Exactly Solvable (QES) systems} pioneered by Turbiner \cite{turbiner1987,turbiner1988,turbiner1989,turbiner1988}, and as a rule they never have essential singularities of the type $e^{-1/g}$. The perturbation theory  in QES for the relevant part of the spectrum systems is convergent. However, such system often have real non-trivial saddles for which it seems impossible to argue that they do not contribute, in contradiction with the absence of contributions of the type $e^{-1/g}$. Further, related set of systems which we call pseudo-QES systems have a completely convergent perturbation theory even though they cannot be solved for exactly. }

In both QES and pseudo-QES systems we could rightfully argue that there is no need of Borel sum since the perturbation series is convergent and therefore well-defined. A trivial example of this situation is already given by the ground state of the SUSY Quantum Mechanics~\cite{Witten:1981nf}, which is zero to all orders of perturbation theory. Because of this the Borel plane is free from singularities, rendering the perturbation theory trivially unambiguous. Hence, one may be tempted to conclude that the perturbative and non-perturbative effects are completely disconnected. While this statement is not in contradiction with resurgence, it would appear that the role of resurgence in these systems is trivial as the different sectors appear to be independent from each other and no cancellation among them is required. We are going to argue that in reality the situation is more subtle, and that resurgence is still governing the interplay between different sectors encoded in the analytical properties of all the various contributions. So, much like the grin of the mythological Cheshire cat, resurgent properties linger even when its main role seems to vanish. For this reason we call this property the Cheshire Cat resurgence\footnote{Many thanks to Thomas Sch\"afer pointing out the analogy.}.

Finally we discuss a remarkable property of the perturbative expansion of the energy levels in these systems: the self-resurgence. Namely because the crucial contributions to the energy is coming from a complex saddle, which generically gives a complex contribution to the energy (except when the hidden topological angle is quantized).  It is well known that large-orders of perturbative expansion around the perturbative vacuum is dictated by the early terms of the perturbation theory around a complex saddle, such as instanton-anti-instanton saddle. 
On the other hand the 
early terms of perturbative corrections \emph{around the complex saddle solution} can be directly connected to the early terms of the perturbative expansion \emph{around the trivial vacuum} via the generalization of Dunne-\"Unsal relation \cite{Dunne:2013ada} to these system, as shown in this work.  By transitivity we are therefore able establish a one-to-one relation between the early terms of the perturbation theory and the late--asymptotic terms \emph{of the same series}.
It is possible that this remarkable property of the perturbation series is connected to the work of Dingle\footnote{We are thankful to M.V.~Berry for sharing the early manuscript with us.} \cite{berry2016} where self-resurgence appears in expansions of functions which are themselves resurgence functions. 

\par{In this work, we address the following questions to shed more light on the resurgent structure of the perturbation theory and to better understand the relevance of the resurgence theory in quantum systems: }

\begin{itemize}
\item \emph{When does an all-order convergent perturbation theory converges? When does it give an exact answer? }
\item \emph{What is the role of the non-perturbative complex saddles and quantization of  hidden topological angles in path integral and  in connection to the  convergence of perturbation theory? }
\end{itemize}
 Addressing these questions allows us to  explore the connections between various approaches to quantum mechanical systems 
\begin{itemize}
\item[{\bf  1)}]  The nature of perturbation theory: convergent {\it vs.} asymptotic, 
\item[{\bf  2)}]  The nature of complex saddles: quantized {\it vs.} unquantized hidden topological angles (associated with the saddles of holomorphized path integrals), 
\item[{\bf  3)}]  Supersymmetry and QES {\it vs.} non-solvability,\footnote{Algebraically non-solvable systems may potentially be solvable in the sense of  a resurgent-transseries.} 
\item[{\bf  4)}]  Resurgence in disguise {\it vs.} explicit resurgence.
\end{itemize}
We demonstrate that  these properties are intimately related: the left and right of the {\it vs.} in four categories are interconnected.  
Quite possibly, these connections transcends quantum mechanics and generalize to QFT,   in particular, we expect that the quantization of the hidden topological angle to imply convergence of perturbation theory for some states, even in QFT, see Section \ref{sec:QFT}.

\par{We study a one-parameter, $\zeta$, family of quantum mechanical systems. Varying $\zeta$ will allow us to interpolate between a purely bosonic theory and quantum mechanical systems with a number of fermions. The integer values of $\zeta$ are particularly interesting since we recover the simplest supersymmetric quantum mechanics when $\zeta=1$, and for other positive integer values of $\zeta$ the lowest $\zeta$ eigenstates are algebraically solvable. As soon as $\zeta$ differs from an integer value, the system ceases to be solvable, and its perturbation series become divergent. For this one-parameter family of quantum systems, the ones with analytic perturbation series consist of a measure-zero subset, and live as limits of generic values of $\zeta$. In other words, the resurgence theory connects the perturbative and non-perturbative sectors and guarantees the well-definiteness of the system for any generic value of $\zeta$. All these relations survive the special values of $\zeta$ as well, which is a Cheshire Cat resurgence.   }

\par{This work also  unifies  various approaches to understand quantum mechanical systems parametrized by $\zeta$: Studies of the perturbation theory via Bender-Wu method\footnote{The mathematica package \texttt{BenderWu} developed in \cite{Sulejmanpasic:2016fwr} was used throughout this work. An up to date version can be freely obtained from the Wolfram package repository at \url{http://library.wolfram.com/infocenter/MathSource/9479/}\;.}~\cite{Bender:1969si,Bender:1973rz,Sulejmanpasic:2016fwr}, the semi-classical analysis and holomorphization of path integral within Picard-Lefschetz theory~\cite{Behtash:2015kna, Behtash:2015loa}, supersymmetric quantum mechanics~\cite{Witten:1982df, Cooper:1994eh} and quasi-exact solvability~\cite{Turbiner:2016aum, Klishevich:2000dp}, and resurgence theory applied to quantum mechanics~\cite{Dunne:2014bca, Dunne:2013ada, Jentschura:2004jg}. In the course of exploring these connection, we also resolve some old standing puzzles in the literature of these topics mentioned above. For various known things, we give new streamlined  arguments. In the remainder of the Introduction we will introduce the two models we study and review the main conclusions of the paper.}

\par{The paper is organized as follows: In the rest of this section we discuss our setup, our main results and the two puzzles related to the QES systems. Sections \ref{sec:DSG} and \ref{sec:TDW} are dedicated to the detailed resolution of the two puzzles, the Cheshire Cat resurgence and the self-resurgence properties in DSG and TDW systems, respectively. In \ref{sec:QFT} we discuss possible connections and parallels with QFT, while in \ref{sec:conclusions} we give conclusions and summary.}

\subsection{The fermions and the $\zeta$-deformed systems}
The main outcome of this work is  most simply described  by considering  the  Euclidean bosonic  Lagrangians of the type,
 \begin{equation}
{\cal L}^E_\zeta=  \frac{1}{g} \left(  {1\over 2} \dot x^2  +   V(x) \right)   \qquad  V(x) =  {1\over 2} \left(W'(x)\right)^2  + {1\over 2} \zeta g  W''(x), 
\label{lag-2}
\end{equation}
where $W(x)$ is auxiliary potential, $V(x)$ is the potential, $g$ is coupling, and $\zeta$ is a deformation parameter whose consequence we explore.  We say that the theory is purely bosonic when $\zeta=0$. The $\zeta=\pm1$ cases are Fermi-Bose sectors (or spin up/down sectors) of supersymmetric quantum mechanics, where $W(x)$ is called the super-potential. We further assume that there are instanton solutions in the purely bosonic theory, but this assumption can be dropped for generalization. 

Quantum mechanics defined by the Lagrangian (\ref{lag-2}) has a formal similarity with some quantum field theories, such as adjoint QCD. 
To give enough motivation, let us consider a quantum mechanical systems with one bosonic and  $N_f$ Grassmann valued fields:
\begin{equation}
{\cal L}^E= \frac{1}{g} \left( {1\over 2}\dot x^2  + {1\over 2} \left(W'(x)\right)^2  \right) 
   + {1\over 2} i (\bar \psi_i \dot \psi_i -   \dot {\bar \psi}_i \psi_i) 
   + {1\over 2} W''(x) [\bar \psi_i, \psi_i ]  \, ,  
   \qquad i=1, \cdots, N_f \, . 
\label{lag-2}
\end{equation}
Because of the fermion flavor symmetry of (\ref{lag-2}), this quantum system is decomposed into superselection sectors defined by fermion number $k$ with degeneracy $N_{f}\choose k$:
\begin{equation} 
H = \bigoplus_{k=0}^{N_f} { H}_{(N_f, k)}    \, . 
\label{Hamk-selec}
\end{equation} 
and the  Hamiltonian for the level $k$ is 
\begin{equation}
{ H}_{(N_f, k)} =  \frac{g}{2}   \frac{  p^2}{2}+   \frac{1}{2g} \left (W'(x)\right)^2  +  {1\over 2}( 2k - N_f)
           W''(x),   \qquad \qquad  k=0,  \cdots, N_f \, . 
\label{Hamk}
\end{equation}
We now find that $\zeta$ in (\ref{lag-2}) is a generalization of $2k-N_f$, and the ${1\over2}\zeta g W''(x)$ term should be viewed as a fermion loop effect. The graded decomposition  \eqref{Hamk-selec} is a generalization of the Bose-Fermi paired Hamiltonians of supersymmetric system.

Although it is not  a priori clear how the ideas around QES systems  (related to integer values  of $\zeta$ in lagrangians in \eqref{lag-2})
generalize to QFT,  the way such systems is presented in \eqref{lag-2} has obvious generalization to QFT. In fact, 
the Lagrangian (\ref{lag-2}) is inspired from multi-flavor quantum  field theory studies.  For example, consider a non-linear sigma model in 2d and add to it a fermionic super-partner.  And then, continue adding $N_f$ fermionic flavors~\cite{Dunne:2012ae, Dunne:2012zk}.   Or similarly, consider adding 
adjoint representation fermions to 4d Yang-Mills, which becomes supersymmetric at $N_f=1$ and some multi-flavor theory for $N_f\geq 2$. There is by now building up evidence that  these QFTs are special in some ways, and carry over some of the interesting aspects of supersymmetric theory~\cite{Unsal:2007jx, Basar:2013sza}, and see Section \ref{sec:QFT} for summary.

The QM systems with Lagrangians \eqref{lag-2} also appear in other contexts. For example, 
\begin{itemize}
\item  Bosonic coordinate $x(t)$ and one Grassmann valued coordinate $\psi(t)$, with a deformation of the Yukawa term by the parameter $\zeta$ \cite{Balitsky:1985in}.
\item Bosonic coordinate $x(t)$, and $W'' (x(t))$ (``magnetic field")  coupled to spin in Bloch representation, via an abelian Berry phase term \cite{Behtash:2015loa}. $\zeta$ acquires an interpretation as analytic continuation of spin quantum number. 
\end{itemize}
See Ref.~\cite{Behtash:2015loa} for a more detailed discussion.

\subsection{The nature of the perturbation theory} 

In general, the perturbation theory in powers of coupling constant $g$ is a divergent asymptotic expansion because of the factorial growth of coefficients. According to resurgence, the asymptotic nature of the perturbation series is caused by the existence of the other saddles of the action
(see, e.g., \cite{Berry657,delabaere2002,Howls} for examples of one-dimensional integrals).  
A way to make sense out of divergent asymptotic expansion is the lateral Borel sum, i.e, directional Laplace integration of the Borel transform. Borel sum assigns a holomorphic function on a Stokes sector to the asymptotic series. At least, in one-dimensional integrals, the geometric realization of Borel resummation is the integration over the Lefschetz thimbles, see for example \cite{Cherman:2014ofa}. This procedure identifies late terms of the perturbation series around the perturbative vacuum with early terms of the asymptotic expansion around another saddle. 

In the case of quantum mechanics, the Planck constant $\hbar$ or the coupling $g$ is an expansion parameter and takes positive values. In many examples, the Borel transform has a singularity on the positive real axis, which causes an imaginary ambiguity in the Borel sum. 
Bogomolny and Zinn-Justin illustrated that the ambiguity is canceled by an instanton--anti-instanton contribution~\cite{Bogomolny:1980ur,ZinnJustin:1981dx}, and this motivates that resurgence works also for quantum mechanics. This was originally demonstrated for the leading asymptotic growth, but its generalization to all orders is given in~\cite{Dunne:2014bca,Dunne:2013ada}. There is a proposal to give a geometric interpretation of this Bogomolny--Zinn-Justin prescription in terms of Lefschetz thimbles, see  the discussion in Section 4  of~\cite{Behtash:2015loa} and~\cite{Misumi:2015dua}.  

In order to tell our story more concretely,  we will use two exemplary Hamiltonian
\be
H=\frac{g}{2}\:p^2+  \frac{1}{2g} \Big( W'(x)^2+\zeta g  W''(x)\Big), 
\label{Ham-zeta}
\ee
with
\begin{alignat}{3}\label{Ws}
  &W(x)=-{\omega}\cos x, &&  \qquad&&{\rm Double \; Sine\; Gordon \; (DSG)} \\\nonumber
  &W(x)= \frac{x^3}{3}- \frac{\omega^2}{4} x, && &&{\rm Tilted \; Double \; Well \; (TDW)}
\end{alignat}
for general values of $\zeta$.  However, we believe our findings generalize to all potentials discussed \cite{Klishevich:2000dp} straightforwardly, as well as to all of the Quasi-Exactly solvable systems \cite{turbiner1987,turbiner1988,turbiner1989,turbiner1988}. 
 \begin{figure}[t] 
   \centering
   \includegraphics[width= 0.80\textwidth]{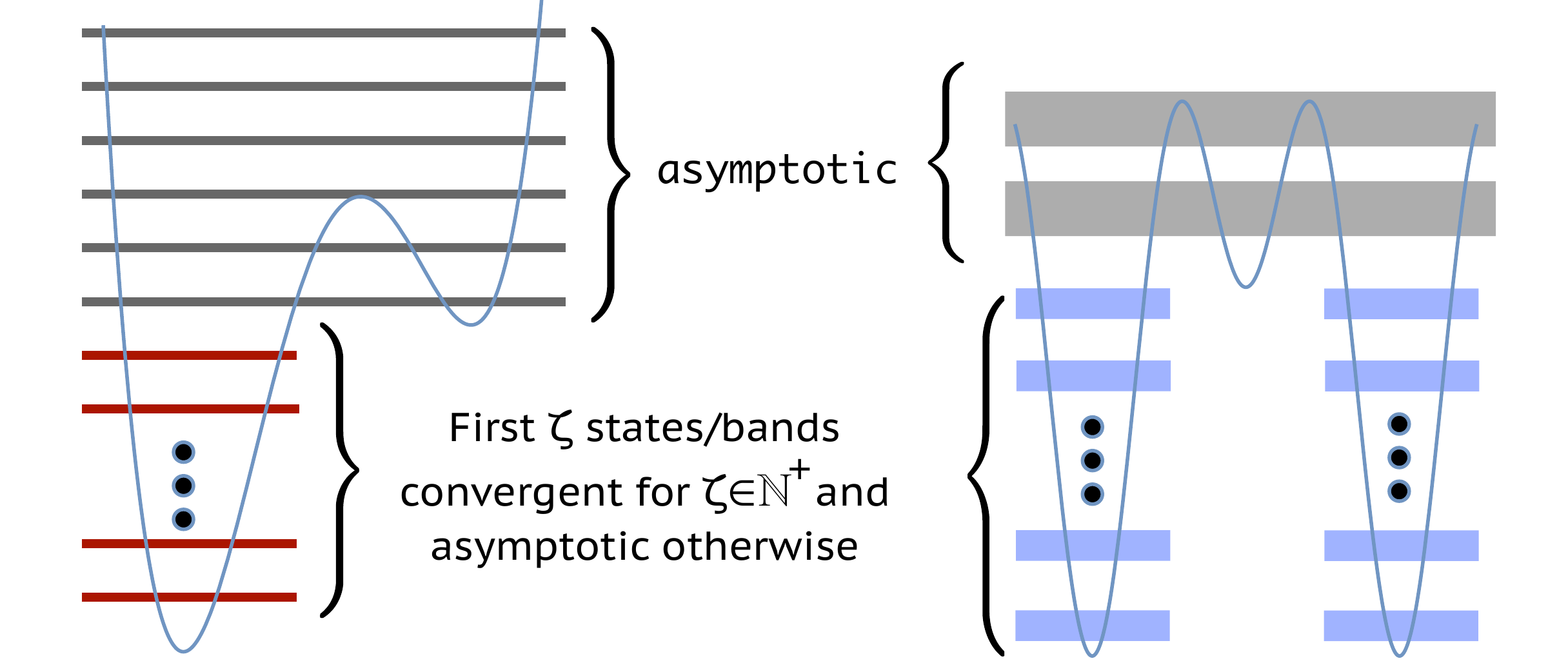} 
   \caption{For integer $\zeta$, perturbation theory for lowest $\zeta$ states is convergent (e.g. $\zeta=1$ is the supersymmetric case), but others are divergent. 
   For non-integer $\zeta$, perturbation theory for all states is asymptotic. 
The energy bands of the DSG system describe dependence of energy levels on the topological $\theta$ angle. 
}
   \label{fig:convergent-asymptotic}
\end{figure}

Let $E^{\rm pert.}(\nu,g,\zeta)$   denote perturbative expansion of the energy for the level number $\nu$ as   ($\nu=0$ is the ground state) 
 \begin{align}
E^{\rm pert.} (\nu,g,\zeta) =  \sum_{n=0}^{\infty}   a_{n}(\nu,\zeta)\, g^n  = a_{0}(\nu,\zeta) + a_{1}(\nu,\zeta) \,g + a_{2}(\nu,\zeta)\, g^2 + \cdots,
\label{pert-exp}
 \end{align}
 By examining the  large-orders of perturbation theory by using the method of Bender and Wu  \cite{Bender:1969si,Bender:1973rz} generalized in \cite{Sulejmanpasic:2016fwr} to arbitrary potentials, 
 we find that the large-order behavior of the expansion coefficients for a level $\nu$ behaves as\footnote{
We determined this growth  in two independent ways. By using complex instanton/bion calculus, and assuming resurgence cancellation of the imaginary parts, this behavior is required. On the other hand  we explicitly computed the perturbative coefficients via the \texttt{BenderWu}  analysis, in excellent agreement with the instanton/bion calculus. The relation of the asymptotic growth to instantons and bions is the reason that the factors ${\cal M},A$ and $S_b$, characterizing the nature of the complex-bion appear. This will be discussed in detail in the rest of the paper.}
 \begin{align}
  a_{n}(\nu,\zeta)  & \sim - \frac{\cal M}{2\pi} \frac{1}{\nu!}  \frac{1}{(2 A^2)^{\zeta -2 \nu -1} } \frac{1}{\Gamma(1 + \nu-\zeta)}  
   \frac{(n -\zeta + 2\nu)!} {(S_b)^{n-\zeta+ 2 \nu + 1}}  \cr
&   \times   \left( b_0(\nu,\zeta) +  \frac{ S_b\,  b_1(\nu,\zeta) }{n-\zeta+ 2 \nu} +    \frac{ S_b^2 \, b_2(\nu,\zeta)}{(n-\zeta+ 2 \nu) (n-\zeta+ 2 \nu-1) }  + \cdots
   \right),
   \label{BW-1}
 \end{align}
where
 \begin{align}
A,{\cal M}&= 2,  \qquad S_b= 2S_I= 2 \times 2 \;= 4 \qquad {\rm DSG}, \cr
A,{\cal M}&= 1,  \qquad S_b= 2S_I= 2 \times \frac{1}{6}=  \frac{1}{3} \qquad {\rm TDW}. 
\label{A-Sb}
 \end{align} 
Here, $S_I$ is the instanton action,    $S_b=2S_I$ is the complex-bion action,   $A$ is the coefficient defined by (note that we set the natural frequency $\omega$ to unity)
 \be
 A=\lim_{t\rightarrow \pm \infty} \dot x_I(t)e^{|t|},
 \ee 
 where $x_I(t)$ is the instanton solution, and $\cal M$ comes from the multiplicity of the complex-bion solution\footnote{In other words, while there is only one complex solution contributing to the double well potential, for the double sine-Gordon there are two complex bion solutions.}.
 At this stage, $b_i(\nu,\zeta)$ just describe correction terms which can in general depend on $\zeta$, but it will connect to perturbation theory around the complex saddles in an interesting way shortly.

The equation  \eqref{BW-1} is an  insightful  formula which deserves multiple comments. It indeed exhibits the generic  $\frac{n!}{(2S_I)^n}$ growth, but there is a curious $\frac{1}{\Gamma(1 + \nu-\zeta)}$ pre-factor which  makes things rather interesting: 
\begin{itemize}
\item[{\bf  1a)}]   For $\zeta \in \mathbb N^{+}$, the leading asymptotic part of the perturbative expansion {\it vanishes} for level numbers $\nu \leq |\zeta-1|$.  By using  an exact Bender-Wu analysis, we also demonstrate that the perturbation theory for those $\zeta$ levels is {\it convergent}. 
The natural question is what is special for this class of theories? 

\item[{\bf  1b)}]     For $\zeta \notin \mathbb N^{+}$,  the perturbation theory for all levels is asymptotic.  They are asymptotic in an expected manner $\sim \frac{n!}{(2S_I)^n}$.  This is the generic behavior.  
\end{itemize}
For  $\zeta=1$, the theory is supersymmetric and  perturbation theory for the ground state   $\nu=0$ is zero, but higher states show asymptotic expansions.  For  $\zeta \in \mathbb N^{+}$ deformed theories, more states have convergent perturbation series. 
See Fig.\ref{fig:convergent-asymptotic} which summarizes this perturbative findings. 

\subsection{The role of complex saddles in the semiclassical analysis}   
In Ref.~\cite{Behtash:2015loa}, it was argued that consistent semi-classical analysis requires the inclusion of complex saddles in the semi-classical expansion. This requires that the real coordinate $x(t)$ to be promoted to the complex coordinate $z(t)= x(t) + i y(t) $ and the path integral to be performed over complex integration cycles passing through the saddles. The saddles are, in general, solutions to the holomorphic Newton's equation in the inverted potential. 
In this way,  complex saddle solutions are found contributing to the ground state energy. Because of their complex nature and their relationship to instanton--anti-instanton, they are called complex bions\footnote{The complex bion is  an exact solution in the $\zeta$-deformed theory.  For small $\zeta g$, It can approximately  be described as an instanton-anti-instanton correlated pair integrated over its quasi-zero mode Lefschetz thimble.  We will use both instanton language and bion language interchangeably (see the discussion in \cite{Behtash:2015loa}).}. 

 The leading non-perturbative contribution of complex bion  $ [{\cal CB}]_{\pm} $ (or equivalently instanton--anti-instanton $[\I \bar \I]_{\pm})$ saddle to the energy  level $\nu$  is given by,
 \begin{align}\nonumber
 E^{\rm n.p.}_{\pm}(\nu,g,\zeta) =  [{\cal CB}]_{\pm} = [\I \bar \I]_{\pm} =  -\frac{1}{2\pi} \frac{\mathcal M}{\nu!}  &\left( \frac{g}{2A^2} \right)^{\zeta- 2 \nu -1} 
    \Gamma(\zeta - \nu)  e^{ \pm i \pi (\zeta - \nu)  } \\
    &\times e^{-S_b/g} \left( b_0(\nu,\zeta) + b_1(\nu,\zeta)\, g + \cdots \right),
    \label{c-b}
 \end{align}
 where $A$ and $S_b$ are defined in \eqref{A-Sb}. The exponent of $e^{ \pm i \pi (\zeta - \nu)  } $ is the phase associated with the complex saddle and its descent manifold, and is called the  \emph{hidden topological angle} (HTA)~\cite{Behtash:2015kna,Behtash:2015loa}. 
The sum  $\sum_{n\in \mathbb{N}} b_{n}(\nu,\zeta)\,g^n \equiv { \cal P}_{\rm fluc}(\nu, g, \zeta) $ denotes perturbative fluctuations around the complex saddle contributions to  level $\nu$.
The HTA of the complex bion solution turns out to be extremely important for resolving some old standing puzzles stemming from the QES solutions. 

The  imaginary  ambiguous parts of the complex bion amplitude can be found by using the reflection formula $\Gamma(\zeta-\nu) {\sin \pi(\zeta-\nu) } = \frac{\pi}{ \Gamma(1-\zeta+\nu)   }$ for the Gamma-function: 
 \begin{align}
{\rm Im}\, E^{\rm n.p.}_{\pm}(\nu,g,\zeta)   &=  \mp \frac{1}{2} \frac{\mathcal M}{\nu!}  \left( \frac{g}{2A^2} \right)^{\zeta- 2 \nu -1} 
   \frac{1} {\Gamma(1+ \nu -\zeta ) } e^{-S_b/g}   \left( b_0(\nu,\zeta) + b_1(\nu,\zeta)\, g + \cdots \right)
   \label{im-cb}
 \end{align}
Just like the Bender-Wu large order result \eqref{BW-1},  there is again  intriguing structure associated with this formula which distinguishes $\zeta \in \mathbb N^{+}$ due to the curious factor $\frac{1}{\Gamma(1 + \nu-\zeta)}$ in \eqref{im-cb}:

\begin{itemize}
\item[{\bf  2a)}]   For $\zeta \in \mathbb N^{+}$,  for which hidden topological angle is quantized, the imaginary ambiguity in the energy disappears.  
What is again special for this class of theories? 
\item[{\bf  2b)}]     For $\zeta \notin \mathbb N^{+}$, the Borel sum of \eqref{BW-1} is ambiguous and has an imaginary ambiguity. This ambiguity must be exactly canceled by the imaginary part of the complex-bion contribution in \eqref{im-cb}, as the energy spectrum must be real. 
\end{itemize}

We also note that there exists a real bion configuration for the DSG system, but there is no such configuration for the TDW.  
The real bion is a real saddle, and hence, it does not possess an HTA. The real bion  contribution to 
 energy  level $\nu$  is given by: 
 \begin{align}\nonumber
 E^{\rm n.p.}(\nu,g,\zeta) =  [{\cal RB}] = [\I  \I] =  -\frac{\mathcal M}{2\pi}  \frac{(-1)^\nu}{\nu!}  &\left( \frac{g}{2A^2} \right)^{\zeta- 2 \nu -1} 
    \Gamma(\zeta - \nu) \\
    &\times e^{-S_b/g} \left( b_0(\nu,\zeta)+ b_1(\nu,\zeta)\, g + \cdots \right)
    \label{r-b}
 \end{align}
Note that for the ground state is, the real bion always reduces the energy, while the for the higher states it alternate as $(-1)^\nu$. Also note that the multiplicity $\mathcal M$ of the real bion is again $\mathcal M=2$, just like that of the complex bion.

\subsection{Supersymmetry and Quasi-Exact Solvability}
Both the quantization of the  hidden topological angle as well as convergence of perturbation theory for   $\zeta \in \mathbb N^{+}$ suggest that there must be something very special about these QM systems. In particular, these systems must realize some generalization of the supersymmetric quantum mechanics. Indeed, this turns out to be the case. 

For either $W(x)$  given in \eqref{Ws} as well as  a very large-class of  $W(x)$ studied in  \cite{Klishevich:2000dp}, we believe that   perturbation theory for the lowest lying $\zeta$ states  is always convergent for   $\zeta \in \mathbb N^{+}$.  The question is whether there is a non-vanishing non-perturbative contribution or not? This is equivalent to the question of dynamical supersymmetry breaking in the $\zeta=1 $ system: 
\begin{itemize}
\item  When  $e^{+ W(x)}$  is   normalizable, the first $\zeta$ states of the $\zeta$-deformed theory are algebraically solvable with the following wave functions, 
\begin{align}
\;\; \Psi_i(x) = P_i(\xi(x))  e^{+ W(x)}, \qquad  i=0, 1, \cdots, \zeta-1, 
\end{align}
where $P_i$ is a set of polynomials in the natural variable $\xi(x)$ of the problem. For $\zeta=1$, this means that supersymmetry is unbroken. 
\item  When  $e^{+ W(x)}$  is   non-normalizable, non-vanishing non-perturbative contribution must exist. The reason is that this solution is generated by the perturbation theory, so it is an exact all-orders perturbative answer. However, since this state does not belong to the Hilbert space due to its non-normalizability, the true energy must be non-perturbatively shifted  to amend it. The situation is entirely parallel in the case the supersymmetric limit when $\zeta=1$, in which case the supersymmetry is dynamically broken by non-perturbative effects \cite{Witten:1981nf}. 
\end{itemize}
In both cases, the perturbation theory for the first $\zeta\in\mathbb N^+$ states converges.

\subsection{Two puzzles of QES}\label{sec:puzzles}
At this point, we wish to point out that 
our work also explains a puzzle emanating from the literature of the QES systems\footnote{We would like to thank Edward Shuryak and Sasha Turbiner for drawing our attention to these puzzles.}: 
\begin{description}
\item[Puzzle 1)] 
For the DSG,  the lowest $\zeta$ states are algebraically solvable. The exact energy expressions which are algebraic in the coupling constant, $g$. At the same time, this system has obvious real instanton type saddles (what we called real bion). This would potentially give a non-algebraic contribution $ e^{-S_{b}/g} $ to the energy. From the exact solutions it can be explicitly seen that no such non-perturbative terms appear. 
\item[Puzzle 2)]
For the TDW,  the lowest $\zeta$ states are not algebraically solvable, these are not QES systems, but their perturbation theory converges. Since the all-order perturbative result does not belong to the Hilbert space (i.e. is not normalizable), there must exist a non-perturbative shift in energy of the form  $ e^{-S_{b}/g} $, but there are no such  real saddles for such a system.
\end{description}
We show that the resolution of these puzzles are given by the realization that apart from the real saddle contributions, there exists another contribution, the \emph{complex bion}:
{\bf Puzzle~1} is solved because real and complex bions {\it exactly} cancel their non-perturbative contributions with each other for the lowest $\zeta$ states; see Section \ref{sec:DSG}. 
{\bf Puzzle~2} is solved because there exist complex saddles contributing to the  energy level $\nu$, and no real non-perturbative saddle to compensate it; see Section \ref{sec:TDW}. 
This provides further evidence that complex paths and saddles are integral to the semi-classical expansion. 

We also find that the convergence of the perturbation theory at these special points in the parameter space is \emph{insufficient} in order to judge whether the perturbation theory gives an exact answer, and that cancellation between contributions of real and complex non-perturbative saddles gives a condition for the perturbation theory being exact. To employ the power of resurgence, such systems must be studied as integer limits of $\zeta$. This type of resurgence we call the Cheshire Cat resurgence, and discuss it next.

 \hspace{.5cm}
 
\noindent\includegraphics[width=1\textwidth]{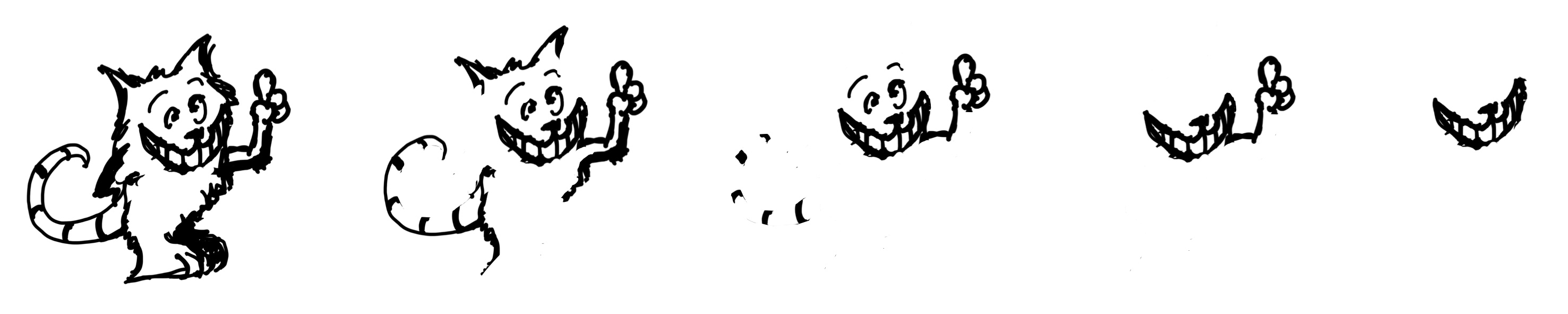}

\subsection[Cheshire Cat resurgence]{Cheshire Cat resurgence\footnote{We are thankful to Roman Sulejmanpasic for the artwork.}}

In our examples, the perturbation series is convergent for $\zeta\in\mathbb{N}^+$.
As convergent series implies no ambiguity, it would seem that the role of resurgence in such systems is trivial as no cancellation between sectors is required, so we could not know whether complex bions contribute. 
We shall nevertheless see that convergence of the perturbation theory no longer holds under a tiny deformation of the theory, such as extension of $\zeta\in\mathbb{N}^+$ to generic $\zeta$. 
Once this is done, the entire structure of resurgence is reestablished. All the relations obtained by resurgence survive even in the limit of convergent perturbation series. We will call it a \emph{Cheshire Cat resurgence}, whose distinguishing features is that from time to time its body disappears,   while its  iconic grin remains. 

\par{For $\zeta \notin \mathbb N^{+}$, by using Bender-Wu analysis, we can do left/right resummation of perturbation theory, and prove that the ambiguity in the Borel sum  ${\cal S}_{\pm }  {E}^{\rm pert.}  (\nu,g,\zeta)  $   
cancels exactly the ambiguity in the complex bion amplitude, $[{\cal CB}]_{\pm}$.  Namely, at leading order, we obtain }
\begin{align} 
\Im\, \Big[   {\cal S}_{\pm }  E^{\rm pert.}  (\nu,g,\zeta)   + [{\cal CB}]_{\pm} (\nu,g,\zeta)   \Big]= 0.
\label{res-cancel}
\end{align}
For $\zeta \in \mathbb N^{+}$, perturbation theory converges, and the complex bion amplitude becomes ambiguity free: 
\begin{align}
 &\Im\,   {\cal S}_{\pm }  E^{\rm pert.} (\nu,g,\zeta)    =   0 \cr
 &\Im\,   [{\cal CB}]_{\pm} (\nu,g,\zeta)     = 0. 
\end{align}
By using this relation, we find that the convergence of the perturbation theory corresponds to the quantization of the hidden topological angle to $\theta_{\mathrm{HTA}} \in \pi (\mathbb Z-0) $.  
In order to extract the non-perturbative information from resurgence at $\zeta\in\mathbb N^+$, let us look more closely at their behaviors as a function of $\zeta$. 
For $\zeta \rightarrow  \mathbb N^{+}$,  we find that the imaginary part of the complex bion amplitude and the  large-orders of perturbation theory behave as  (for example, for ground state, setting  $\nu=0$  in   \eqref{im-cb} and  \eqref{BW-1})
 \begin{align}
{\rm Im}\, E^{\rm n.p.}_{\pm} (\nu =0,g,\zeta)  &=  \mp \frac{1}{2}  \left( \frac{g}{2A^2} \right)^{\zeta -1} 
   \frac{1} {\Gamma(1 -\zeta ) } e^{-S_b/g}   \left( b_0(0,\zeta) + b_1(0,\zeta)\, g + b_2(0,\zeta)\, g^2 +  \cdots \right), \cr  
  a_{n}(\nu=0,\zeta) & =- \frac{{\cal M}}{2\pi}   \frac{1}{(2 A^2)^{\zeta  -1} } \frac{1}{\Gamma(1 -\zeta)}  
   \frac{(n -\zeta )!} {(S_b)^{n-\zeta + 1}}\nonumber\\&\times \left( b_0(0,\zeta) +  \frac{ (S_b)  b_1(0,\zeta) }{n-\zeta} +   \frac{ (S_b)^2  b_2(0,\zeta) }{(n-\zeta)(n-\zeta-1)} +    
   \cdots  
   \right).    
\label{cat}
 \end{align}
 Despite the fact that both expressions become zero in the   $\zeta \rightarrow  \mathbb N^{+}$  limit due to overall  $ \frac{1} {\Gamma(1 -\zeta ) } $ factor,   and the resurgent cancellation  seems to disappear,  the footprint of resurgence is still present in the theory.   This may also be viewed as an analyticity in $\zeta$; if resurgent cancellation works infinitesimal away from $\zeta \in \mathbb N^{+}$, its remnant must be present even in the limit. 

By employing the Cheshire Cat resurgence we can justify our claim that the complex bion gives a contribution to the semiclassical analysis even when $\zeta\in\mathbb{N}^+$. This claim is the essential ingredient to solve the puzzles in QES literatures. 

\subsection{Self-resurgence and the Dunne-\"Unsal relation}

In this section we discuss another remarkable feature of the systems we study. Namely since in both systems we study the ambiguity of non-perturbative contribution is given by a complex saddle point--- the complex bion in the cases we study--- the perturbative corrections to this saddle will have a one-to-one correspondence with the corrections to the leading asymptotic growth of the perturbation theory. Although we always keep in mind the two systems we study (i.e. the DSG and the TDW), it is worth noting that the arguments we present in here are generally applicable to any system  for which a complex saddle contributes to the energy shift, and for which the Dunne-\"Unsal relation holds. 

To show the self-resurgent properties, note that the coefficients $b_i$ of the large order expansion \eqref{BW-1} correspond to the perturbative corrections around the complex bion solution via the resurgent cancellation \eqref{res-cancel}, i.e.
\be
\text{Im }\, E^{\rm n.p.}_\pm (\nu,g,\zeta) = (\dots)e^{-S_b/g}\text{Im}(e^{\pm i\zeta\pi})\mathcal P_{\rm fluc}(\nu,g,\zeta)\;.
\ee
where $\mathcal P_{\rm fluc}(\nu,g,\zeta)$ is the perturbative expansion around the complex saddle, normalized so that $\mathcal P_{\rm fluc}(\nu,0,\zeta)=1$. We write a formal expansion of this object as
\be
\mathcal P_{\rm fluc}(\nu,g,\zeta)=\sum_{i=0}^\infty b_{i}(\nu,\zeta)\,g^i\;,
\ee
where $b_0=1$. On the one hand the form of of the large order growth \eqref{BW-1} is fixed by the requirement that the ambiguity resulting from the complex bion is cancelled by the Borel summation of the perturbation theory. On the other hand, Dunne and one of us (M\"U) showed \cite{Dunne:2014bca} that the perturbation around non-trivial saddles can be related to the perturbation theory around the trivial vacuum in a constructive way. By writing an analogous formula for the systems we study, we are able to relate the perturbative expansion around the trivial vacuum to the expansion around \emph{the complex saddle}. These two facts then seem to imply that the perturbation theory around it both dictates and is dictated by (respectively) the late and early terms of the perturbation theory around the trivial vacuum. But if this is the case it means that the early terms of the perturbative series, ``echoing'' on the non-perturbative ``mountain'', dictate late terms of \emph{of the same series}. For this reason it is appropriate to call this phenomenon \emph{echo-resurgence} or \emph{self-resurgence}. 

Let us see in more detail how this works. The formal power-expansion of the energy in coupling $g$ of the energy level $\nu$ is given by
\begin{align}
E^{\rm pert.} (\nu,g,\zeta)&=  a_{0}(\nu,\zeta) + a_{1}(\nu,\zeta) g+ a_{2}(\nu,\zeta) g^2+ \cdots
\label{perturb-00}
    \end{align}

Now the natural generalization of the Dunne-\"Unsal relation for an arbitrary complex bion is 
\be
\mathcal P_{\rm fluc}(\nu,g,\zeta)= \frac{\partial{E^{\rm pert.}}}{\partial\nu}  \exp\left[ S_b  \int_{0}^{g} \frac{dg}{g^2} \left(  \frac{\partial{E^{\rm
pert.}}}{\partial \nu}  - a'_0(\nu,\zeta)-a_1'(\nu,\zeta)g \right) \right].  
\ee
Writing 
\begin{align}
\frac{\partial E^{\rm pert.}}{\partial \nu}  &=  \sum_{n=0}^{\infty}
 a_n'(\nu,\zeta)\, g^n,
\label{perturb-00}
    \end{align}
    where the prime indicates differentiation with respect to $\nu$.
We get that
    \begin{align}
{ \cal P}_{\rm fluc}(\nu, g,\zeta) &= \left(1+  \sum_{n=1}^{\infty}  a_n'(\nu,\zeta)\,  g^n \right) \, e^{ S_b
\sum_{n=1}^{\infty}  \frac{ 1}{n}a_{n+1}'(\nu,\zeta)\, g^n  } \\\nonumber
&= (1+  a_1'(\nu,\zeta)\,  g + a_2'(\nu,\zeta)\,  g^2 +  a_3'(\nu,\zeta)\, g^3 + \cdots ) \\&\hspace{3cm}\times \exp\left[
S_b \left( a_2'(\nu,\zeta)\,  g + \frac{ a_3' (\nu,\zeta)\,g^2}{2} +    \frac{ a_4'(\nu,\zeta)\,
g^3}{3}   + \cdots  \right)\right].  \cr
\label{magic-formula-00}
\end{align} 
 Equivalently, we can write $b_{i}(\nu,\zeta)$'s in terms of derivatives of $a_{i}(\nu,\zeta)$'s,
      \begin{align}\nonumber
b_0(\nu,\zeta)&=1, \\\nonumber
b_1(\nu,\zeta) &= a_1'(\nu,\zeta) + S_{b} \, a_2'(\nu,\zeta),  \\\nonumber
b_2(\nu,\zeta)&=  a_2'(\nu,\zeta)  + S_{b} \,a_1'(\nu,\zeta) a_2'(\nu,\zeta)  + \frac{1}{2} S_{b}^2\, a_2'(\nu,\zeta)^2 +
\frac{1}{2} S_{b}\,  a_3'(\nu,\zeta),\\\nonumber
b_3(\nu,\zeta)&=a_3'(\nu,\zeta)+S_b\left(a_2'(\nu,\zeta)^2+\frac{1}{2} a_1'(\nu,\zeta) a_3(\nu,\zeta)+\frac{1}{3}a_{4}'(\nu,\zeta)\right)\\&+\frac{1}{2}S_b^2\left(a_1'(\nu,\zeta) a_2'(\nu,\zeta)^2+a_2'(\nu,\zeta) a_3'(\nu,\zeta)\right)+\frac{1}{6}S_{b}^{3}\,a_2'(\nu,\zeta)^3.
\end{align}

By plugging in \eqref{BW-1} we get that the large order corrections of the perturbative expansion are given by
\begin{align}
 a_{n}(\nu,\zeta)&= - \frac{\cal M}{2\pi} \frac{1}{\nu!}  \frac{1}{(2 A^2)^{\zeta -2 \nu -1} } \frac{1}{\Gamma(1 + \nu-\zeta)}  
   \frac{(n -\zeta + 2\nu)!} {(S_b)^{n-\zeta+ 2 \nu + 1}} \\\nonumber
   &\times\left[1  +\frac{ S_b  \,  ( a_1'(\nu,\zeta) + S_{b}\,  a_2' (\nu,\zeta) ) } {n-\zeta}+ \right. \\ &\hspace{2cm}+\left.\frac{S_b^2\,
  \left( a_2'(\nu,\zeta)  + S_{b}\, a_1'(\nu,\zeta) a_2'(\nu,\zeta)  + \frac{1}{2} S_{b}^2\, a_2'(\nu,\zeta)^2 +
\frac{1}{2} S_{b}\,  a_3'(\nu,\zeta)  \right)}{ (n-\zeta)(n-\zeta-1)}+ \cdots \right],
 \label{BW-P-4}  
\end{align}
what we have obtain is nothing short of remarkable! Indeed the expression above relates the asymptotic coefficients of the perturbation theory $a_n$, for $n\gg1$, to the derivatives $\partial_\nu a_n(\nu,\zeta)=a_n'(\nu,\zeta)$ for $n=1,2,3,\cdots$. For this reason we say that if the Dunne-\"{U}nsal relation holds for a system in which complex saddles contribute, the perturbation series of the energy is said to be \emph{self-resurgent}.

We take an opportunity now to comment on the possible interpretation of this self-resurgence formula as being related to Dingle's self-resurgence formula (see M.V.~Berry \cite{berry2016}) which is a general property of resurgent functions which are themselves functions of resurgent functions. Namely it is likely\footnote{TS would like to thank M.V.~Berry for drawing our attention to this possibility.} that the self-resurgent properties of the systems we study imply that the energy is not simply a resurgent function of two independent arguments $\nu$ and $g$, but that they are related in some way. Indeed in \cite{Dunne:2014bca}, the energy is written as $E(\nu(g),g)$, where part of the dependence on $g$ is placed into a functional dependence on $\nu$. On the other hand, here we obtained the self-resurgent formula by the utilization of the Dunne-\"Unsal relation, a formula which is only known for systems who's WKB Riemann sheet is topologically a torus \cite{Basar:2015xna}. If by virtue of \cite{berry2016} the self-resurgent property is a general property of eigenvalue problems, this may give insight into what the generalization of the Dunne-\"Unsal relation for higher genus WKB Reimann surfaces is.


\section{Resolving Puzzle 1: The Double Sine-Gordon system}\label{sec:DSG}
In this section, we provide the resolution of the \textbf{Puzzle 1} that is described in Section \ref{sec:puzzles}. A subset of the lowest energy eigenstates of the Double Sine-Gordon (DSG) are exactly solvable, and the corresponding energy eigenvalues are known to be algebraic in coupling constant $g$. On the other hand, according to the textbook semi-classical analysis, the system posses real instantons, what we call real bions (because these are really correlated two-instanton events with a characteristic size parametrically 
larger than an instanton.) They should introduce non-algebraic $e^{-S_{b}/g}$ contributions to the energy eigenvalues. The presence of complex bions, in addition to the real bions, lies in the heart of the solution to the apparent discrepancy. The complex bion contribution cancels precisely the one coming from real bions. The concrete relation of this QM system with quantum field theory is described in 
Section \ref{sec:QFT}.

Here we consider the DSG system and analyze it in detail. The Hamiltonian is
\begin{align}
\label{eq:DSGpot}
&H=-\frac{g}{2}\frac{\partial^2}{\partial x^2}+\frac{1}{2g}\Big(W'(x)^2-\zeta g W''(x)\Big),\\
&W(x)=-\omega\cos x\;, \cr
&V(x)=\frac{\omega^2}{2g}\sin^2x-\zeta\frac{\omega}{2}\cos x, 
\end{align}
where $\omega$ is the curvature at $x=0$ to leading order in the expansion parameter $g$ and $\zeta$ is an a priori free parameter.
The Schr\"odinger equation reads
\be
-g\frac{\psi''(x)}{2}+V(x)\psi(x)=E\psi(x)\;.
\ee
Since the potential is periodic, the wave-function can have Bloch periodicity $\psi(x+2\pi)=\psi(x) e^{i\theta}$. By changing the $\theta$-angle we can scan the band of the potential.
Below, we examine this class of potential by using the methods outlined in the Introduction for general values of $\zeta$.

\begin{figure}[t] 
   \centering
   \includegraphics[width=\textwidth]{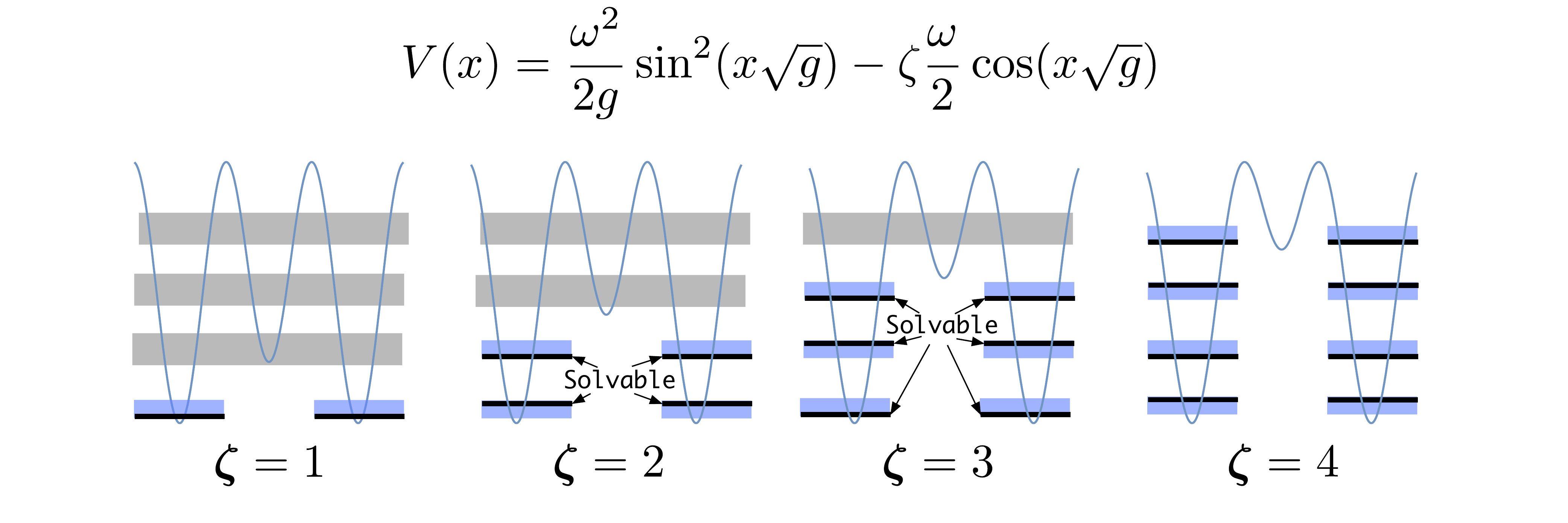} 
   \caption{An illustration of the exactly solvable states. The blue-shaded rectangles represent bands by changing the theta angle, who's width is non-perturbative and not exactly solvable for any $\zeta$. However, it is possible to solve either for the energy of the top or of the bottom of the band when $\zeta$ is an odd or an even integer respectively. Note that $\zeta=1$ case is a supersymmetric limit, and the bottom of the band corresponds to the supersymmetric ground state given by $\psi_0=e^{-W(x)}=e^{\frac{\omega}{g}\cos(x\sqrt g)}$.}
   \label{fig:bands}
\end{figure}

First, recall that for the $\zeta=1$ case the above system reduces to the well known case of supersymmetric quantum mechanics, with the ground state energy $E_0=0$ and the ground state wave-function
\be
\psi_0=e^{-W(x)/g}=e^{\frac{\omega}{g}\cos(x\sqrt g)}\;
\label{susy-wf}
\ee
This solution determines the bottom of the lowest-lying band, or the ground state at $\theta=0$.
What is much less appreciated is that for  any $\zeta\in \mathbb N^+$,  
it is always possible to find either the bottom or the top of the first $\zeta$ bands  (see Fig. \ref{fig:bands}) analytically. The method which allows one to determine these edges of the band goes under the name of \emph{Quasi-Exact Solvability} (QES).  We discuss this next. 

\subsection{Quasi-Exact Solvability for $\zeta \in \mathbb N^{+}$}
 The definition of the QES is that  a finite part of the spectrum is algebraically exactly solvable (see \cite{Turbiner:2016aum} for a recent review of QES). 
Let us denote by $\mathcal{H}_0$ the finite-dimensional subspace spanned by those eigenstates, which are algebraically solved under a certain boundary condition. 
 
The method of QES relies on rewriting the eigenvalue problem by a suitably chosen Ansatz of the wave-function 
\begin{align}
\psi(x)=u(x)e^{-W(x)/g}\;.
\end{align}
The Schr\"odinger equation for $\psi$ then turns into the eigenvalue equation for $u(x)$, of the form
\be
\hat hu=Eu,
\ee
where $\hat h$ is a second order differential operator  given by
\be
\hat h=\omega\left[-\frac{g}{2\omega}\frac{d^2}{dx^2}+ \sin x \frac{d}{dx}-\frac{\zeta-1}{2} \cos x\right].
\label{ham-su2}
\ee
Note that we can set $\omega=1$. To reinstate it in the result we simply need to replace the coupling $g\rightarrow g/\omega$ and the energy $E\rightarrow \omega E$.

In order to turn the Hamiltonian operator into a matrix eigenvalue equation in the subspace $\mathcal H_0$,  it is useful 
to introduce differential operators which form a representation of $SU(2) $ algebra, 
\be
J_+=e^{-ix}\left(j-i\frac{d}{dx}\right)\;,\qquad J_-=e^{ix}\left(j+i\frac{d}{dx}\right)\;,\qquad J_3= i\frac{d}{dx}, 
\label{generators}
\ee
with a Casimir $(J_+J_-+J_-J_+)/2+J_3^2=j(j+1)$. The eigenfunctions of $J_3$ are $u_m=Ne^{-imx}$, with $m=-j,-j+1,\dots, j$ and they form a multiplet in the $2j+1$ dimensional representation of $SU(2)$.  $\mathcal H_{0}$ is the span of $u_m$, and exact solutions will be decomposable within this subspace. 

The Hamiltonian   $\hat h$   given in \eqref{ham-su2} can be expressed in terms of generators \eqref{generators}
\be\label{eq:htoop}
\hat h= \frac{g}{2}J_3^2-\frac{1}{2}(J_++J_-)=-\frac{g}{2}\frac{d^2}{dx^2}+\sin x\, \frac{d}{dx}-j\cos x,
\ee
provided  we identify $j=\frac{\zeta-1}{2}$. Since $j$ must be a non-negative integer or half-integer, we see that $\zeta$ must be a positive integer.

Now note that there exists an abstract scalar product invariant under the action of the $SU(2)$ group in question\footnote{Note that $J_\pm$ are not Hermitian conjugates of each other under the naive $L^2$ norm. We can, instead, introduce the invariant norm under the $SU(2)$ group, so that it automatically makes the generators $J_a$ invariant as it must follow that
$$
(u,v)=(Uu,Uv)= (u,v)+t^a(iJ^a u,v)+t^a(u,iJ^a v)+o((t^a)^2)\Rightarrow  (J^a u,v)=(u,J^a v)\;,
$$
where wrote $U=e^{iJ^a t^a}$, where $J^a$ are generators of $SU(2)$, and $t^a$ are parameters of the transformation. } and under which the states $u_m$ are orthogonal, i.e. $(u_m,u_s)=\delta_{ms}$ (see \cite{vilenkin}). To determine the norm of $u_m$ we note that the action of $J_\pm$ is given by
\be
J_\pm u_m=\sqrt{(j\mp m)(j\pm m+1)}u_{m\pm 1}\;.
\ee
Choosing $u_{-j}\equiv e^{ijx}$, $(u_{-j},u_{-j})\equiv 1$ by definition, we can construct all $u_m$ by a successive action of $J_+$. This gives
\be
u_m=\sqrt\frac{(2j)!}{(j-m)! (j+m)!}\;e^{-imx}\;, \qquad m=-j,-j+1,\dots, j
\ee

This fact naturally splits the Hilbert space $\cal H$ into a subspace invariant under the $\hat h$ and the rest. Note that this $SU(2)$ group is not a symmetry of the underlying theory, i.e. Hamiltonian is \emph{not} invariant under the action of this $SU(2)$. This is clear from the form of the operator $\hat h$, given by \eqref{eq:htoop}, which is clearly not $SU(2)$ invariant. Rather the space spanned by $u_m$ is invariant under the action of $\hat h$, which allows for an algebraic solution of one part of the spectrum. Indeed in this subspace the operator $\hat h$ attains a (tridiagonal) matrix form
\be
\scriptsize
\hat h_0=\frac{1}{2}\begin{pmatrix}
g(-j)^2&-\sqrt{2j}&0&0&\dots&&&&\\\\
-\sqrt{2j}&g(-j+1)^2&-\sqrt{2(2j-1)}&0&\dots&&&&\\
0&-\sqrt{2(2j-1)}&g(-j+2)^2&\ddots&\\
0&0&\ddots &\ddots&&&\\
\vdots&\vdots & & &  gm^2&-\sqrt{(j-m)(j+m+1)}\\\\
& & & &  -\sqrt{(j-m)(j+m+1)}&g(m+1)^2&\\
& & & &  & \ddots&\\
& & & &  & &g(j-1)^2&-\sqrt{2j}\\
& & & &  & &-\sqrt{2j}&gj^2\\
\end{pmatrix},
\ee
where the subscript on $\hat h$ implies the restriction of the total Hilbert space to the subspace $\mathcal{H}_0$ invariant under the action of the $SU(2)$ group generated by $J_3, J_\pm$. As we will only be concerned by this subspace, we will drop the subscript $0$ in what follows.

\subsubsection{Exact solutions for $\zeta=1,2,3$, and $4$}
Let us explicitly consider  $\mathcal H_0$ and  $\hat h$  for the cases $\zeta=1,2,3,4$. 
\subsubsection*{$\zeta=1$ (Supersymmetric) case: One exactly solvable state}

This is the supersymmetric case, and it evidently requires  $j=0$,  so that the only solvable state is $u(x)=$const. 
This  is precisely the supersymmetric ground state. 
The Hamiltonian action on  ${\mathcal H_0}$  is 
\be  
\hat h=  0,
\ee 
with eigenvalues and eigenfunctions
\be
E_0= 0, \qquad \psi_0= e^{\frac{1}{g} \cos x}. 
\ee
Notice that the wave-function is periodic, so it corresponds to the Bloch angle $\theta=0$.

\subsubsection*{$\zeta=2$ case: Two exactly solvable states}

In this case $j=1/2$ and we have the Hamiltonian acting on ${\mathcal H_0}$ 
\be
\hat h= 
\left(
\begin{array}{cc}
 \frac{g}{8} & -\frac{1}{2} \\
 -\frac{1}{2} & \frac{g}{8} \\
\end{array}
\right),
\ee
with eigenvalues and eigenfunctions
\begin{align}
& \textstyle{ E_0=   - \frac{1}{2} + \frac{g}{8} ; \qquad   \psi_0= \cos(\frac{ x}{2})  \; e^{\frac{1}{g} \cos  x }},
\cr  
& \textstyle{ E_1=   + \frac{1}{2} + \frac{g}{8} ; \qquad   \psi_1= \sin(\frac{ x}{2})  \; e^{\frac{1}{g} \cos  x }}.
\end{align}
Notice that these wave-functions obey anti-periodic boundary condition, so that the Bloch angle is given by $\theta=\pi$. 

\subsubsection*{$\zeta=3$ case: Three exactly solvable states }

If $\zeta=3$ then $j=1$. The Hamiltonian is: 
\be
\hat h= 
\left(
\begin{array}{ccc}
 \frac{g}{2} & -\frac{1}{\sqrt{2}} & 0 \\
 -\frac{1}{\sqrt{2}} & 0 & -\frac{1}{\sqrt{2}} \\
 0 & -\frac{1}{\sqrt{2}} & \frac{g}{2} \\
\end{array}
\right),
\ee
with eigenvalues and eigenfunctions
\begin{align}
& \textstyle{ E_0=  \frac{1}{4}\left(g -  \sqrt{g^2+16}\right)      ; \qquad   \psi_0=    \left(2\cos x+\frac{g + \sqrt{g^2+16}}{2}\right) e^{\frac{1}{g}\cos  x}}, \cr
& \textstyle{ E_1=  \frac{g}{2}      ;  \qquad \qquad  \qquad \qquad \qquad    \psi_1=  \Big( \sin x   \Big) \;e^{\frac{1}{g}\cos  x}}, \cr
& \textstyle{ E_2=  \frac{1}{4}\left(g +  \sqrt{g^2+16}\right)      ; \qquad   \psi_2=    \left(2\cos x+\frac{g - \sqrt{g^2+16}}{2}\right) e^{\frac{1}{g}\cos  x}}.
\label{sol-zeta3}
\end{align}
Just like in the case when $\zeta=1$, the all the wave-functions we found are periodic, therefore the Bloch angle is $\theta=0$ again.

\subsubsection*{$\zeta=4$ case: Four exactly solvable states}
If $\zeta=4$ then $j=3/2$. The Hamiltonian is: 
\begin{align}
\hat h=  \left(
\begin{array}{cccc}
 \frac{9 g}{8} & -\frac{\sqrt{3}}{2} & 0 & 0 \\
 -\frac{\sqrt{3}}{2} & \frac{g}{8} & -1 & 0 \\
 0 & -1 & \frac{g}{8} & -\frac{\sqrt{3}}{2} \\
 0 & 0 & -\frac{\sqrt{3}}{2} & \frac{9 g}{8} \\
\end{array}
\right),
\end{align}
with eigenvalues and eigenfunctions
\begin{align}
& \textstyle{ E_0=-\frac{1}{2}+\frac{5}{8}g- \frac{\sqrt{4+2g+g^2}}{2}; \qquad   \psi_0=  \left( \cos (\frac{3 x}{2}) + \frac{\sqrt{g^2+2 g+4}+g+1}{\sqrt{3}}  \cos(\frac{ x}{2}) \right) \;e^{\frac{1}{g} \cos  x }},
\cr  
&\textstyle{ E_1=+\frac{1}{2}+\frac{5}{8}g- \frac{\sqrt{4-2g+g^2}}{2}; \qquad   \psi_1=  \left( \sin (\frac{3 x}{2}) + \frac{\sqrt{g^2-2 g+4}+g-1}{\sqrt{3}}  \sin(\frac{ x}{2}) \right) \;e^{\frac{1}{g} \cos  x } },     \cr 
& \textstyle{ E_2=-\frac{1}{2}+\frac{5}{8}g+ \frac{\sqrt{4+2g+g^2}}{2}   ; \qquad   \psi_2=  \left( \cos (\frac{3 x}{2}) + \frac{ - \sqrt{g^2+2 g+4}+g+1}{\sqrt{3}}  \cos(\frac{ x}{2}) \right) \;e^{\frac{1}{g} \cos  x }}, \cr
&\textstyle{ E_3= +\frac{1}{2}+\frac{5}{8}g+ \frac{\sqrt{4-2g+g^2}}{2};  \qquad   \psi_3=  \left( \sin (\frac{3 x}{2}) + \frac{ -\sqrt{g^2-2 g+4}+g-1}{\sqrt{3}}  \sin(\frac{ x}{2}) \right) \;e^{\frac{1}{g} \cos  x } }.
\end{align}
As in the case of $\zeta=2$, wave-functions obey the anti-periodic boundary condition.

\subsubsection{General $\zeta \in \mathbb N^{+}$ case and Ince-Polynomials}
For general $\zeta \in \mathbb N^{+}$ theory, the first $\zeta$ level are algebraically solvable, corresponding to 
$j= \frac{\zeta-1}{2}$ representation of $SU(2)$.  
 The $2j+1 = \zeta$ solutions are of the form 
\be
\psi_i ( x)=  P^{(\zeta-1)}_i ( \cos( x/2),\sin( x/2)) e^{\frac{1}{g}\cos  x} , \qquad i=0, 1, \cdots, \zeta-1
\ee
where $P^{(\zeta-1)}_i ( \cos( x/2),\sin( x/2))$  is an $(\zeta-1)^{\rm th}$ order polynomial with trigonometric arguments, $ \cos( x/2)$ and $\sin( x/2)$. These are called {\it Ince-polynomials} (see, e.g., Sec.~28.31 of Ref.~\cite{NIST:DLMF}). 
The wave functions for the exactly soluble subset obey the boundary conditions:
 \begin{align}
 \psi_i (  x + 2\pi) = (-1)^{\zeta-1}  \;  \psi_i (  x)
\label{exact-bc}
 \end{align}
 \begin{itemize}

\item \noindent {\bf $\bm \zeta$-odd:} In this case, the exactly solvable subset obey   periodic boundary conditions \eqref{exact-bc}.  See 
Fig.~\ref{fig:bands}.  This  corresponds to topological  theta angle zero, $\theta=0$. Note that for the $\nu=0$ band, this corresponds to the bottom of the band,  while top of the band  corresponds to $\theta=\pi$. The bottom of the $\nu=1$  band also corresponds to $\theta=\pi$ and is again not algebraically solvable, but the top of the $\nu=1$ band correspond to $\theta=0$ and is algebraically solvable. This pattern continues for all odd-$\zeta$ values. 
 
 \item \noindent {\bf $\bm \zeta$-even:} In this case, the exactly solvable subset obey   anti-periodic boundary conditions \eqref{exact-bc}.  See 
Fig.~\ref{fig:bands}.  This  corresponds to setting topological  theta angle to  $\theta=\pi$. Note that for the $\nu=0$ band, this corresponds to the top of the band.  Note that the bottom of the $\nu=1$  band also corresponds to $\theta=\pi$ and it is also 
 algebraically solvable.  The rest of neither bands is exactly solvable. 
This pattern continues for all even-$\zeta$ values. 
\end{itemize}

\subsection{Complex saddles and the role of the hidden topological  angle}
\label{sec:puzzle}
There is a long-standing puzzle in the literature of the QES systems. 
Because of its algebraic nature, the exact solutions in QES systems have no non-perturbative contributions of the form $e^{-S_b/g}$.  On the other hand, there is a real non-perturbative saddle in the DSG system, which we refer to as the real bion $[\cal {RB}]$ \cite{Behtash:2015loa}. This real saddle interpolates from $ x=0$  to $ x=2\pi$, which may be interpreted as the exact version of instanton-instanton  $[\I\I]$ correlated event. There is no reason why this object would not contribute to the semiclassical analysis.  Indeed such paths must be present, as only they couple to the Bloch $\theta$ angle. If such paths did not contribute, the Bloch bands would not exists. Why does this contribution disappear in the exact energy expression? To our best knowledge, the resolution of this puzzle is not known in the QES literature. 

The explanation of the above mentioned puzzle is very similar to that of the instanton--anti-instanton contributions in a supersymmetric theory \cite{Behtash:2015loa}.  In other words, for the case of $\zeta=1$ supersymmetric theory, the vacuum energy is zero to all orders in the perturbation theory. Contributions of the real bion renders the ground-state energy negative at $\theta=0$, which clashes with the supersymmetry algebra. 
In \cite{Behtash:2015loa}, a complex multivalued saddle is found and called the complex bion, whose contribution to vacuum energy is positive,  $\Delta E^{cb} \sim - e^{\pm i \pi}  e^{-S_b/g} $, and it cancels the real-bion contribution   $\Delta E^{rb} \sim -   e^{-S_b/g} $   exactly. 
This is the first hint for building up Picard-Lefschetz theory for path integrals\footnote{As Picard-Lefshetz theory of path integrals is not a complete theory, we should clarify what we mean by this.  By Picard-Lefshetz theory we mean a meaningful and systematic expansion of the observables which have a path-integral representation into contributions coming from various saddles of the action.  Note that we do not a priori refer to the nature, structure and construction of Lefshetz thimbles. More concretely, we do not claim that the dual of the Lefschetz thimble associated with the complex bion has nonzero intersection number with the original integration cycle.   Our intention is to build this theory for path integrals, and in this work, by using Bender-Wu analysis, resurgence, QES, supersymmetry  and complexified path integral, we take mileage in this direction, and provide an almost complete picture for the systems we study.  }, because it demonstrates that the complex saddles of this type  must be included in the semiclassical expansion.  In the $\zeta$-deformed theories, we find a  similar  exact cancellation mechanisms for $\zeta\in\mathbb{N}^+$.

 \begin{itemize}
\item  \noindent {\bf $\bm \zeta$-odd:}  As asserted above, the lowest $\zeta$ states are exactly solvable at $\theta=0$, and thus there must exist an exact cancellation between real 
 and complex bion saddles. Indeed, we find
  \begin{align}
 E^{\rm n.p.}(\nu,g,\zeta) &= 2 [{\cal RB}] + 2[{\cal CB}]_{\pm}   \cr
&    =2 \left(- (-1)^\nu -  e^{ \pm i \pi (\zeta - \nu) }\right)_{\zeta=1,3, \cdots }  e^{-S_b/g}   \cdots   \cr
&=2 (-1)^{\nu+1} (1+ e^{i \pi \zeta})_{\zeta=1,3, \cdots }  e^{-S_b/g}   =0  
    \label{rb-cb}
 \end{align}
 For level $\nu=0,2, \cdots, \zeta-2$, the real bion reduces the energy, while the complex bion increases it and the two cancel exactly. 
 For level $\nu=1,3, \cdots, \zeta-1 $, the real bion increases  the energy, while the complex bion reduces  it and the two cancel exactly.  
 The cancellation between the two is a consequence of the destructive interference induced by {\it hidden topological angle $\theta_{\mathrm{HTA}}= \zeta \pi$} associated with the complex saddle. 
  \item    \noindent {\bf $\bm \zeta$-even:}  In this case, the lowest $\zeta$ states are exactly solvable at $\theta=\pi$, so we must consider the effect of the topological $\theta$ angle for the QES. 
Since the real and complex bions have the winding numbers $1$ and $0$, respectively, we find that
\begin{align}
 [{\cal RB}] (\theta)=  [{\cal RB}]  e^{i \theta}, \qquad  [{\cal CB}] (\theta)=  [{\cal CB}]  
    \label{turn-theta}
 \end{align}
 As a result of this, the contribution of real and complex bion to the energy level $\nu$ for the  case of  even $\zeta$ takes the form  
  \begin{align}
 E^{\rm n.p.}(\nu,g,\zeta) &=  ([{\cal RB}] +c.c.)+2 [{\cal CB}]  \cr
&    = 2\left(- (-1)^\nu \cos\theta|_{\theta=\pi}  -  e^{ \pm i \pi (\zeta - \nu) }\right)_{\zeta=2,4, \cdots }  e^{-S_b/g}   \cdots   \cr
&= 2 (-1)^{\nu+1} ( \cos\theta|_{\theta=\pi} + e^{i \pi \zeta})_{\zeta=2,4, \cdots }e^{-S_b/g}  \cdots   =0  
    \label{rb-cb-2}
 \end{align}
 For level $\nu=0,2, \cdots, \zeta-1$, the real bion increases the energy, while the complex bion reduces it and the two cancel exactly. 
 For level $\nu=1,3, \cdots, \zeta-2 $, the real bion reduces  the energy, while the complex bion increases  it and the two cancel exactly.  
 The cancellation between the two is a consequence of the destructive interference induced also by {\it ordinary  topological angle}  $\theta$ associated with the  real saddle. 
\end{itemize} 

We find the mechanism described in \eqref{rb-cb} and \eqref{rb-cb-2} nothing short of remarkable. It is due to this exact non-perturbative cancellation mechanisms induced by the interplay of the hidden topological angle with the ordinary topological angle is necessary for the exact algebraic solvability of the states in ${\cal H}_0$ in these QES-systems. 

So far, we have shown that a consistent semiclassical picture is given for QES if we take into account the effect of complex bions. Remainder of Section \ref{sec:DSG} is dedicated to the analytic properties off the integer values of $\zeta$. As we shall see, such theories hold much more information about QES systems then would naively be thought.


\subsection{The general $\zeta$-deformed theory}
In this and the next sections, we shall show that the complex bion must be taken into account for the semiclassical analysis by using a resurgence relation. 
For $\zeta\in\mathbb{N}^+$ the perturbation series has a finite convergence radius, and there seems to be no room for resurgence to play into the game. 
In order to understand what is happening better, we are going to compute the perturbation series of the DSG system for a generic $\zeta\in\mathbb{R}$ and establish the intricate relation of the perturbative sector and the complex bion. By using the continuity in the limit $\zeta\to1,2,3,\cdots$, we argue that the complex bion still describes a nonperturbative contribution in the semiclassical analysis without the factorial growth of the perturbation series; we name it a Cheshire Cat effect. 

To compute the perturbation series, we apply the Bender-Wu method~\cite{Bender:1969si, Bender:1973rz, Sulejmanpasic:2016fwr}. 
The Bender-Wu method  is an   algorithm  to compute the high order correction  for an arbitrary energy level in  perturbation theory. 
The main idea of this algorithm is to construct the recursive relation for the perturbative coefficients of the eigen-energy $E$ and eigenfunction $\psi$. 

We demonstrate two remarkable aspect of the perturbation theory. 
\begin{itemize}
\item For $\zeta\in\mathbb{N}^+$, the perturbation theory of the DSG system for $\nu=0,1, \cdots, \zeta-1$ is convergent and exact. For higher energy levels, the perturbation theory yields an asymptotic expansion.  
\item   For   generic $\zeta$, the perturbation theory is always asymptotic.  
\end{itemize}

\subsubsection{Exactly solvablity  from perturbation theory for $\zeta \in \mathbb N^{+}$ }

For $\zeta  \in \mathbb N^{+}$, Bender-Wu equation gives  a convergent results  for the energy   levels  $\nu=0, 1, 2, \cdots, \zeta-1$, and  it gives divergent asymptotic series for level number $\nu \geq \zeta $.    See for example Tables~\ref{DSG1}, \ref{DSG2}, \ref{DSG3}.

For $\zeta=1$, the system is supersymmetric. Indeed, the ground state  ($\nu=0$) energy is zero to all orders in perturbation theory.   For the wave-function,   perturbation theory does not yield zero, but a convergent and exact result for level number $\nu=0$. 
For higher states  $\nu=1,2, \cdots$    in the supersymmetric theory, perturbation theory  is asymptotic. 

As an example of a  convergent (and non-truncating)  perturbation theory,  see  Tables~\ref{DSG3}, let us show the series for the the ground state energy of the $\zeta=3$ system. 
\begin{align}
E^{{\rm pert.}} (\nu=0, g,\zeta)=-1+\frac{g}{4}-\frac{g^2}{32}+\frac{g^4}{2048}-\frac{g^6}{65536}+\frac{5
   g^8}{8388608}-\frac{7 g^{10}}{268435456}+\cdots,
   \end{align}
which is exactly the expansion of $E_0$ in \eqref{sol-zeta3}.

\subsubsection{Asymptotic corrections from the Bender-Wu analysis}
\label{sec:cor}
Studying the Bender-Wu recursion relation,   one can find  the large-order  behavior of  perturbation theory.
\begin{multline}
  a_{n}(\nu,\zeta)   \approx - \frac{1}{\pi} \frac{1}{\nu!}  \frac{1}{8^{\zeta -2 \nu -1} } \frac{1}{\Gamma(1 + \nu-\zeta)}  
   \frac{(n -\zeta + 2\nu)!} {(S_b)^{n-\zeta+ 2 \nu + 1}}  \cr
   \times   \Bigg(1 +  \frac{ S_b\,  b_1(\nu,\zeta)  }{n-\zeta+ 2 \nu} +    \frac{ S_b^2\,  b_2(\nu,\zeta) }{(n-\zeta+ 2 \nu) (n-\zeta+ 2 \nu-1) }\\ + \cdots +\frac{S_b^K\, b_K(\nu,\zeta) }{(n-\zeta+2\nu)(n-\zeta+2\nu-1)\cdots (n-\zeta+2\nu-K)}
   \Bigg),
 \end{multline}
where we set $b_0=1$ and terminated the $1/n$ correction to some finite order $K$. 
This can be done for any state, but here we report for state $\nu=0$ and $\nu=1$. 
We can then use the \texttt{BenderWu} package \cite{Sulejmanpasic:2016fwr} to compute the coefficients $a_{n}(\nu,\zeta)$ to some high order $n=N$, retaining the analytic dependence on $\zeta$. Explicit values of $n=N-K,N-K+1,\dots,N$ can then be plugged into the above approximate equation, giving $K$ equations with $K$ unknowns $b_1(\nu,\zeta) ,b_2(\nu,\zeta) ,\cdots, b_K(\nu,\zeta) $. Taking $K=10$, and expanding in a series in $\zeta$ for $\nu=0$ we get the following numerical values
\begin{align}
b_1(\nu=0,\zeta)&=-0.6249999999802+0.624999999937376\zeta-0.1249999999149091\zeta^2+o(\zeta^310^{-11}),\nonumber\\
b_2(\nu=0,\zeta)&=-0.10156250436+ 0.015625013847 \zeta + 0.117187481117609 \zeta^2\nonumber\\
                        &\hspace{0cm} - 0.0624999852717885 \zeta^3 + 0.00781249267805392 \zeta^4 + o(\zeta^510^{-9}),\nonumber\\
b_3(\nu=0,\zeta)&=-0.116211 + 0.124022 \zeta - 0.00325335 \zeta^2 - 
 0.016603 \zeta^3 - 0.003254481026038194 \zeta^4 \nonumber\\&\hspace{0cm}
 +0.002929444295569087 \zeta^5 - 0.0003254657952886725 \zeta^6 +
 o(\zeta^7 10^{-9}).
 \label{eq:BW_b_coeff_nu0}
\end{align}
We repeat the same for $\nu=1$ and obtain
\begin{align}
b_1(\nu=1,\zeta)&=-2.6249999967918 + 1.374999993327027 \zeta - 
 0.1249999939430032 \zeta^2+o(\zeta^310^{-9}),\nonumber\\
b_2(\nu=1,\zeta)&=1.24218678107 - 1.953123499260 \zeta + 0.929686131557870 \zeta^2\nonumber\\
                        &\hspace{0cm} +0.1562492802762927 \zeta^3 + 0.00781225722624248 \zeta^4+o(\zeta^510^{-8}).,\nonumber\\
b_3(\nu=1,\zeta)&=-0.471608 + 0.473483 \zeta - 0.444525 \zeta^2 + 
 0.278248 \zeta^3 - 0.0774493776897379 \zeta^4 \nonumber\\&\hspace{0cm}
+0.00878340297785316 \zeta^5 - 0.0003246026044364228 \zeta^6 +
 o(\zeta^7 10^{-7}).
 \label{eq:BW_b_coeff_nu1}
\end{align}

The fact that the perturbative coefficients follow the factorial growth given by \eqref{BW-1} suggests that the complex bion must contribute in the semiclassical analysis. By using the continuity in $\zeta$, complex bion gives the contribution also for $\zeta\in\mathbb{N}^+$, which solves the puzzle in QES literature as we discussed in Sec.~\ref{sec:puzzle}. Furthermore, our detailed computation on $b_i(\nu,\zeta) $ gives the conjecture about the perturbative fluctuations around the saddle of complex bions.

\subsection{Self-resurgence and the Dunne-\"Unsal relation} 
\label{CR}
By using Bender-Wu recursion relations, we can derive a perturbative  expansion for the energy eigenvalues $E^{\rm pert.}(\nu, g, \zeta)$  as a function of coupling $g$, level number $\nu$, and parameter $\zeta$. For example, up to fourth order in $g$, we obtain an expression 
\begin{align}
 E^{\rm pert.}(\nu, g, \zeta) &=a_{0}(\nu,\zeta)+a_{1}(\nu,\zeta)\,g+a_{2}(\nu,\zeta)\,g^2+a_{3}(\nu,\zeta)\,g^3+a_{4}(\nu,\zeta)\,g^4+{\cal O}(g^{5})\,,
\end{align}
where
\begin{align}\nonumber
a_{0}(\nu,\zeta)&=\nu +\frac{1}{2}-\frac{\zeta }{2},\\\nonumber
a_{1}(\nu,\zeta)&=\frac{1}{8} \left(2 \zeta  \nu +\zeta -2 \nu ^2-2 \nu -1\right)\\\nonumber
a_{2}(\nu,\zeta)&=\frac{1}{64} \Big(\zeta ^2 (-(2 \nu +1))+\zeta  \left(6 \nu ^2+6 \nu
   +3\right)-2 \left(2 \nu ^3+3 \nu ^2+3 \nu +1\right)\Big),\\\nonumber
a_{3}(\nu,\zeta)&=\frac{1}{256} \Big(\zeta ^3 (2 \nu +1)-6 \zeta ^2 \left(2 \nu ^2+2 \nu
   +1\right)+\zeta  \left(20 \nu ^3+30 \nu ^2+32 \nu +11\right) \Big.\\\nonumber
   &\Big.-2 \left(5 \nu^4+10 \nu ^3+16 \nu ^2+11 \nu +3\right)\Big),\\\nonumber
a_{4}(\nu,\zeta)&=\frac{1}{4096}\Big(-5 \zeta ^4 (2 \nu +1)+48 \zeta ^3 \left(2 \nu ^2+2 \nu +1\right)-2 \zeta
   ^2 \left(142 \nu ^3+213 \nu ^2+233 \nu +81\right)\Big.\\ 
   &+15 \zeta  \left(22 \nu
   ^4+44 \nu ^3+74 \nu ^2+52 \nu +15\right)-2 \left(66 \nu ^5+165 \nu ^4+370 \nu
   ^3+390 \nu ^2+225 \nu +53\right)
\end{align}
As stated earlier, the traditional resurgence connects large-order growth around the perturbative vacuum of perturbation theory to early terms around the instanton--anti-instanton saddle. However, a new type of resurgence,  which follows from exact quantization condition implemented via uniform WKB  approach, offers a constructive version of resurgence \cite{Dunne:2014bca}. It is an early term--early term relation. The knowledge of perturbative expansion around the perturbative saddle at order $g^{n}$ as a function of energy levels is sufficient to deduce the  fluctuations around the 
the leading non-perturbative saddle at order $g^{n -1}$.  
 The non-perturbative  
contribution to the energy for level $\nu$ is
\begin{align}
E^{\rm n.p.}_{\pm} (\nu, g, \zeta) &=  [{\cal RB}] + [{\cal CB}]_{\pm}  \\
& =   -\frac{1}{2\pi}  \frac{2}{\nu!}  \left( \frac{g}{8} \right)^{\zeta- 2 \nu -1}    \Gamma(\zeta - \nu) ( (-1)^\nu +   e^{ \pm i \pi (\zeta - \nu) })      e^{-S_b/g}  
 { \cal P}_{\rm fluc}(\nu, g, \zeta)   \qquad \nonumber
\end{align}  
where ${ \cal P}_{\rm fluc}(\nu,g,\zeta)$ is the fluctuation operator around the real and complex saddle.  We remind the reader that, according to the result of \cite{Dunne:2013ada,Dunne:2014bca} (see also \cite{Gahramanov:2015yxk}) in the case of $\zeta=0$, the fluctuations around an instanton-saddle are completely determined from the perturbative expansions around the trivial saddle. Inspired by this, we give a conjectured form of the relation between $ { \cal P}_{\rm fluc}(\nu,g, \zeta)$  of complex bion and the trivial perturbation theory $E^{\rm pert.} (\nu,g, \zeta)$
\begin{align}
\label{magic-2}
{ \cal P}_{\rm fluc}(\nu, g, \zeta) &=  \frac{\partial{E^{\rm pert.}}}{\partial \nu}  \exp\left[ S_b  \int_0^g  \frac{dg}{g^2} \left(  \frac{\partial{E^{\rm pert.}}}{\partial \nu}  -  1 +  \frac{2g (\nu+ \frac{1}{2} - \frac{\zeta}{2}) }{S_b}   \right) \right].   
\end{align}

How can we check this formula? One way is to show consistency with the exact quantization condition, similar in spirit to the  
Zinn-Justin and Jentshura \cite{Jentschura:2004jg,ZinnJustin:2004ib, ZinnJustin:2004cg}.  We defer the discussion of exact quantization condition  for $\zeta$-deformed theories elsewhere. Instead, from above expression we can identify the $\zeta$-polynomials $b_i(\nu,\zeta)$ by noting that
\begin{align}
&{ \cal P}_{\rm fluc}(\nu, g, \zeta)  = b_0(\nu, \zeta) +   b_1(\nu, \zeta)    g  + b_2(\nu, \zeta)     g^2  
+ \cdots,   \qquad
\label{fluc-3}
\end{align}
so that
\begin{align}
b_0(\nu,\zeta)&=1\;,\cr
b_1(\nu,\zeta)&=\frac{1}{8} \Big(-\zeta ^2+\zeta  (6 \nu +5)-2 \nu  (3 \nu +5)-5\Big), \cr
b_2(\nu,\zeta)&= \frac{1}{128} \Big(\zeta ^4-4 \zeta ^3 (3 \nu +2)+\zeta ^2 \left(48 \nu ^2+56 \nu +15\right)\cr &\hspace{0cm}-2 \zeta 
   \left(36 \nu ^3+60 \nu ^2 +30 \nu -1\right)+36 \nu ^4+80 \nu ^3+60 \nu ^2-4 \nu -13\Big),\cr
b_3(\nu,\zeta)&=\frac{1}{3072}\Big(-\zeta ^6+9 \zeta ^5 (2 \nu +1)-2 \zeta ^4 \left(63 \nu ^2+51 \nu +5\right)+\zeta ^3
   \left(432 \nu ^3+432 \nu ^2+42 \nu -51\right)\cr
&\hspace{0.0cm}-2 \zeta ^2 \left(378 \nu ^4+444 \nu
   ^3+21 \nu ^2-165 \nu +5\right)+3 \zeta  \left(216 \nu ^5+300 \nu ^4-228 \nu ^2+70 \nu
   +127\right)\cr
&\hspace{0cm}-3 \left(72 \nu ^6+120 \nu ^5-152 \nu ^3+70 \nu ^2+254 \nu +119\right)\Big). 
\end{align}
Setting $\nu=0$ and $1$, we can compare them with an estimate to these polynomials in \eqref{eq:BW_b_coeff_nu0} and \eqref{eq:BW_b_coeff_nu1}, respectively. Indeed the reader is welcome to check that the coefficients of \eqref{eq:BW_b_coeff_nu0} and \eqref{eq:BW_b_coeff_nu1} differ from the ones above by no more than 0.06\%. 
This consistency again strengthens the evidence that the complex bion gives a physical contribution for the DSG system in the semiclassical analysis and justifies the Cheshire Cat resurgence. 
For the ground state, set level number $\nu=0$ we obtain 
\begin{align}\label{eq:DU_b_coeff_nu0}
 b_0(\nu=0,\zeta)\Bigg|_{\text{D\"U}} &=1,  \cr
 b_1(\nu=0,\zeta)\Big|_{\text{D\"U}}  &= \frac{1}{8}  \Big(      -5 +   5  \zeta -  \zeta^2   \Big),  \cr
  b_2(\nu=0,\zeta)\Big|_{\text{D\"U}}  &=   \frac{1}{128} \Big(-13  + 2  \zeta + 15  \zeta^2  -8  \zeta^3 +   \zeta^4
\Big),  \cr
  b_3(\nu=0,\zeta)\Big|_{\text{D\"U}}  &=   \frac{1}{3072} \Big( -\zeta ^6+9 \zeta ^5-10 \zeta ^4-51 \zeta ^3-10 \zeta ^2+381 \zeta -357 \Big),
\end{align}
where we have explicitly indicated that the result was obtained from the Dunne-\"Unsal relation \eqref{magic-2}. This confirms, at least to the precision indicated above that the formula \eqref{magic-2} holds.

Let us do the same with $\nu=1$. From the Dunne-\"Unsal relation we have
\begin{align}
&b_0(\nu=1,\zeta)\Big|_{\text{D\"U}}=1,\cr
&b_1(\nu=1,\zeta)\Big|_{\text{D\"U}}=\frac{1}{8} \left(-\zeta ^2+11 \zeta -21\right),\cr
&b_2(\nu=1,\zeta)\Big|_{\text{D\"U}}=\frac{1}{128} \left(\zeta ^4-20 \zeta ^3+119 \zeta ^2-250 \zeta +159\right),\cr
&b_3(\nu=1,\zeta)\Big|_{\text{D\"U}}=\frac{1}{3072}\left(-\zeta ^6+27 \zeta ^5-238 \zeta ^4+855 \zeta ^3-1366 \zeta ^2+1455 \zeta
   -1449\right).
\end{align}
Comparing with \eqref{eq:BW_b_coeff_nu1} we find that the coefficients agree with the above formula to the precision of no more than 0.3\% (most coefficients are below 0.06\%).

\section{Resolving Puzzle 2: Tilted Double-Well}\label{sec:TDW}

In this section, we present the resolution of \textbf{Puzzle 2} of Section \ref{sec:puzzles}. The Tilted Double-Well (TDW) is not a QES system, but the perturbation series converges both for the wave function and energy eigenvalue. The wave function obtained in this manner  is non-normalizable, and therefore, it cannot be a non-perturbative solution. This is again in contrast with the existing textbook semi-classical approach since the TDW potential  does not possess real non-perturbative saddles. 
 In the case of TDW, the complex bions come to rescue too and explain why all orders perturbative solution is not exact.
 
Let us repeat our analysis for the $\zeta$-deformation of the double-well system to get more insight on the connection of the perturbation theory and complex saddles. 
The Hamiltonian takes the same form~\eqref{Ham-zeta}, where the auxiliary potential (or super-potential for $\zeta=1$)  is given by
 \be
 W(x)=\frac{x^3}{3}-\frac{\omega^2 x}{4},
 \ee
 where $\omega$ is the natural frequency of the system.
For simplicity in the remainder of this section we set $\omega=1$. We can always reinstate it by the following replacement $x\rightarrow \omega x$, $g\rightarrow g/\omega^3$ and the energy eigenvalues $E\rightarrow \omega E$.

Further the  system also has   a convergent perturbation series in powers of $g$ for  $\zeta \in\mathbb N^{+}$.  Moreover,  the series sums  to a finite, but incorrect (or rather incomplete) result.  We will derive this result analytically using techniques of QES. 
 For generic $\zeta$, perturbation theory is asymptotic.  
 
Note that subsection \ref{P-QES} should be considered as a review material as it is already discussed in literature in great depth~\cite{Aoyama:1998nt, Aoyama:2001ca}. 
 Here, we briefly discuss it for completeness. The relation of the fluctuations around the complex bion and perturbation theory around the trivial saddle (the Dunne-\"Unsal relation) and the self-resurgence properties of the perturbation theory are new.

\subsection{Pseudo-QES}
\label{P-QES}
We will now apply QES techniques to ``solve'' the TDW problem.  We emphasize again, that this \emph{not} a genuine solution to the full non-perturbative  problem. It provides the all-order perturbative solution correctly but lacks some non-perturbative contributions. 

We start, as usual, with an Ansatz 
\begin{align}
\psi(x)=u(x)e^{W(x)/g},
\label{non-normalizable}
\end{align}
motivated by the supersymmetric case $\zeta=1$. This 
is  a non-normalizable solution to Schr\"odinger  equation, hence not a state in the Hilbert space.  
 The equation for $u(x)$ is given by
\be
-\frac{g}{2}u''(x)-u'(x) W'(x)+\frac{1}{2}(\zeta -1) u(x) W''(x)=Eu(x)\;.
\ee
Plugging in $W'(x)=x^2-1/4$, we obtain the equation
\be
-\frac{1}{2} g\; u''(x)+\left (\frac{1}{4}-x^2\right) u'(x)+x (\zeta -1)\: u(x)=Eu(x)\;.
\ee
Hence we define the reduced hamiltonian
\be
\hat h=-\frac{g}{2}\frac{d^2}{dx^2}+\left(\frac{1}{4}-x^2\right)\frac{d}{dx}+x(\zeta-1)\;.
\ee
The objective now is to find eigenvalues $E$ of this differential operator. 

Now, observe that the operators
\be
J_+=2jx-x^2\frac{d}{dx}\;,\qquad J_-=\frac{d}{dx}\;, \qquad J_3=x\frac{d}{dx}-j\;.
\ee
obey the $SU(2)$ algebra, which for $2j\in \mathbb N^0$ leaves invariant the vector space spanned by polynomials $u_m=N_m x^{j+m}$ for $m=-j,-j+1,\dots, j$. Further, it takes little to check that  $h_T$ can be written as
\be
\hat h=-\frac{g}{2}J_-^2+\left(\frac{1}{2} J_-+J_+\right)+(\zeta-1-2j)x\;.
\ee
Since $2j\in \mathbb N^0$, then choosing $\zeta=2j+1$ allows us to eliminate the last term above, and write $h_T$ entirely in terms of operators $J_\pm,J_3$. The $h_T$ operator acting on the $SU(2)$ invariant subspace spanned  by polynomials $u_m$ is therefore given by 
\be
\hat h=-\frac{g}{2}J_-^2+(J_-/4+J_+)\;,\qquad 2j=\zeta-1\;.
\ee
Let us now solve several specific cases.

\subsubsection{$\zeta=1,2,3,4$ perturbatively exact solutions}
We  consider  $\mathcal H_0$ and  $\hat {h_T}$  for the cases $\zeta=1,2,3,4$. 

\subsubsection*{$\zeta=1$ case (SUSY)}
If $\zeta=1$, then $j=0$ and the only state in the invariant subspace is $u_0=\text{const}$. 
 The Hamiltonian action  on ${\mathcal H_0}$  is 
\be  
\hat {h}=  0 
\ee 
with eigenvalues and eigenfunctions
\be
E(\nu=0)= 0, 
\ee
This is indeed the result of the SUSY system, as $e^{W(x)/g}$ solves the Sch\"odinger equation with energy zero. This state is not normalizable, hence, supersymmetry is broken dynamically. Indeed, a non-perturbative ground state  energy has the form $E_0^{\rm n.p.} \sim e^{{-S_b}/{g}}$. 

\subsubsection*{$\zeta=2$ case}
In this case we have to solve for eigenvalues of the matrix
\be
\hat{h}=\left(
\begin{array}{cc}
 0 &\frac{1}{4} \\
 1 & 0 \\
\end{array}
\right)
\ee
which are simply
\be
E(\nu=0)=-\frac{1}{2}, \qquad E(\nu=1)=\frac{1}{2}\;.
\ee

\subsubsection*{$\zeta=3$ case}
\begin{figure}[tbp] 
   \centering
  $\text{Re }E(g)$\raisebox{-.55\height}{\includegraphics[width=.75\textwidth]{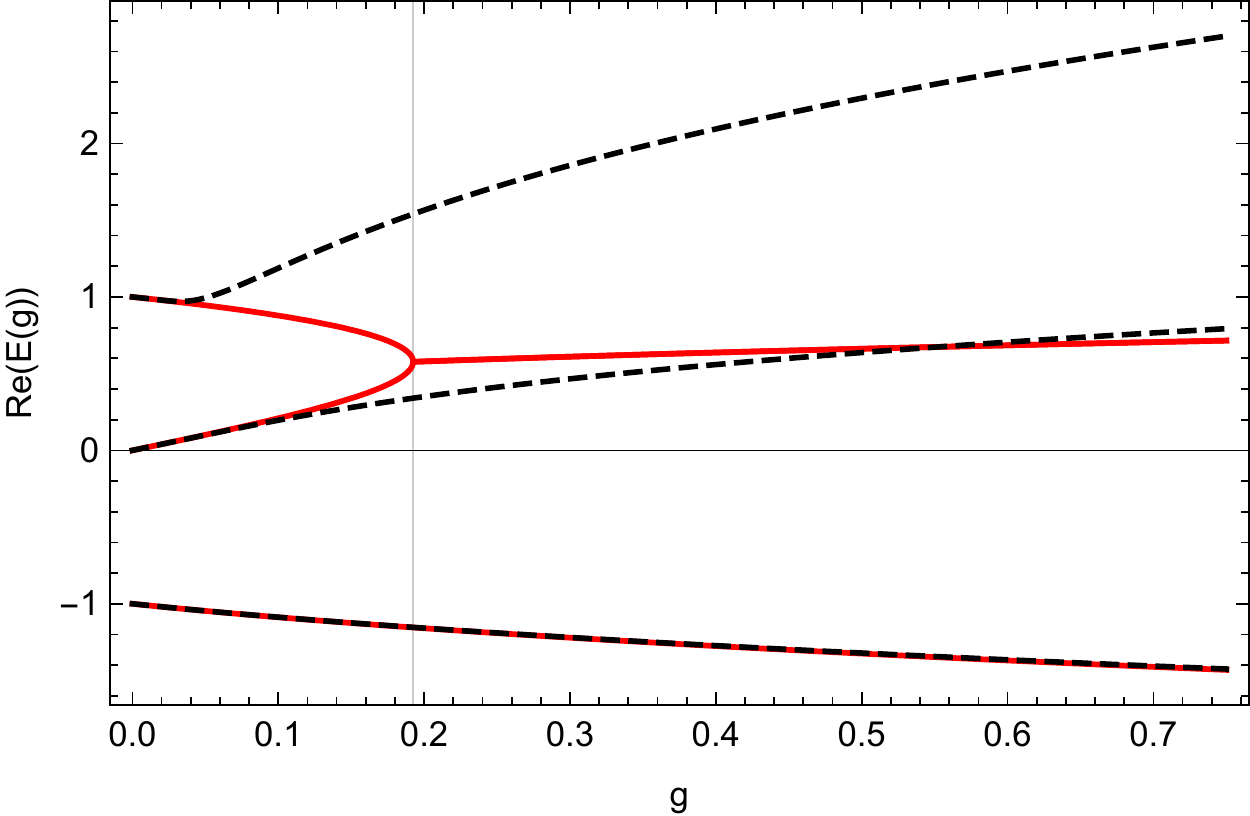} }
   
       \hspace{2cm}{\Large g} 
   \caption{A plot of eigenvalues $E(\nu=\{0,1,2\},g)$ for $\zeta=3$ as a function of coupling $g$. The solid lines represent the real part of the all orders in perturbation theory result \eqref{eq:zeta3_eigenvalues}, while the dashed lines represent the numerical solution to the Schr\"odinger equation. Notice that $E(\nu=1,g)$ and $E(\nu=2,g)$ in \eqref{eq:zeta3_eigenvalues} collide and turn into complex conjugate pairs when $g=\frac{1}{3\sqrt{3}}$. }
   \label{fig:zeta3_eigenvalues}
\end{figure}

Now the matrix becomes
\be
\hat{h}=\left(
\begin{array}{ccc}
 0 & \frac{1}{2\sqrt{2}} & -g \\
 \sqrt{2} & 0 & \frac{1}{2\sqrt{2}} \\
 0 & \sqrt{2} & 0 \\
\end{array}
\right)
\ee
with eigenvalues
\begin{subequations}\label{eq:zeta3_eigenvalues}
\begin{align}
&E(\nu=0,g)=-\frac{2}{\sqrt3} \cos\left[\frac{1}{3}\arccos\left[{3\sqrt3 g}\right]\right]\\
&E(\nu=1,g)=\frac{2}{\sqrt3} \sin\left[\frac{1}{3}\arcsin\left[{3\sqrt3 g}\right]\right]\\
&E(\nu=2,g)=\frac{2}{\sqrt3} \cos\left[\frac{1}{3}\arccos\left[-{3\sqrt3 g}\right]\right]
\end{align}
\end{subequations}
The plot of the real part of $E(\nu=\{0,1,2\},g)$ is given in Fig. \ref{fig:zeta3_eigenvalues}, along with the numerical solution of the Schr\"odinger equation. Notice that while the ground state is described extremely well by the all-orders perturbative result\footnote{It can be shown that the difference is in precise agreement with a complex bion for small $g$.}, $E(\nu=1,g)$ starts deviating significantly already at the coupling $g\approx 0.1$, while $E(\nu=2,g)$ shows a drastic deviation  already at $g\approx 0.05$. Further when $g=\frac{1}{3\sqrt 3}$, the two pseudo-eigenvalues $E(\nu=1,g)$, $E(\nu=2,g)$ merge and for $g>\frac{1}{3\sqrt 3}$ they become complex and turn into each other's complex conjugate pairs. This of course cannot happen for actual eigenvalues of the Schr\"odinger equation.

 \subsubsection*{$\zeta=4$ case}
  \begin{figure}[tbp] 
   \centering
   $\text{Re }E(g)$\raisebox{-.55\height}{\includegraphics[width=.75\textwidth]{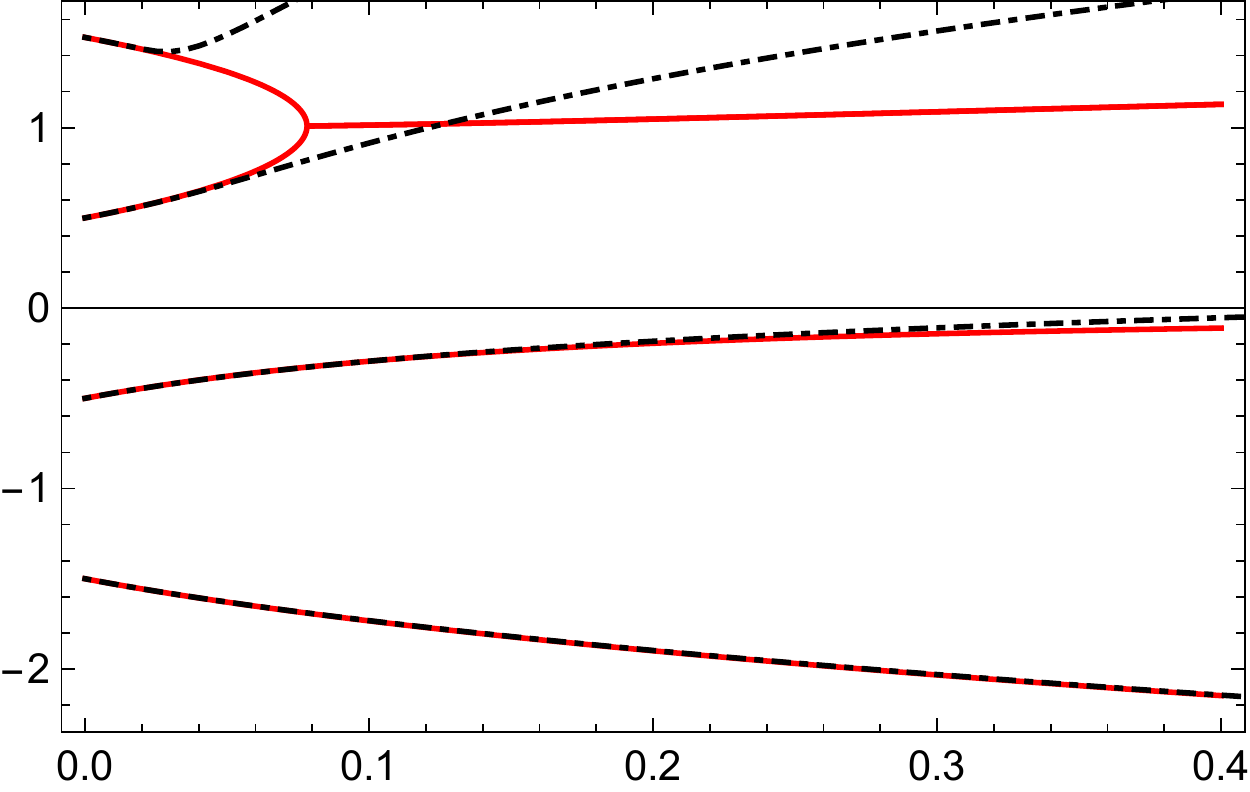}}
   
      \hspace{2cm}{\Large g} 
   \caption{A plot of perturbative eigenvalues (solid line) and the numerical solutions (dashed line) for the states $E(\nu=\{0,1,2,3\},g)$ for $\zeta=4$ as a function of coupling $g$ which are solutions of \eqref{eq:zeta4_eigenvalues}. Notice that while  $E(\nu=\{0,1\},g)$ agree quite well (up to non-perturbative corrections), the perturbative values of $E(\nu=2,g)$ and $E(\nu=3,g)$ merge at some value of $g$ and turn into complex conjugate pairs.}
   \label{fig:zeta4_eigenvalues}
\end{figure}
 The $\hat h$ matrix is given by
 \be
\hat h= \left(
\begin{array}{cccc}
 0 & \frac{\sqrt{3}}{4} & -\sqrt{3} g & 0 \\
 \sqrt{3} & 0 & \frac{1}{2} & -\sqrt{3} g \\
 0 & 2 & 0 & \frac{\sqrt{3}}{4} \\
 0 & 0 & \sqrt{3} & 0 \\
\end{array}
\right)
 \ee
 with the characteristic equation
 \be\label{eq:zeta4_eigenvalues}
\left(E^2-\frac{1}{4}\right)\left(E^2-\frac{9}{4}\right)= -12 g E 
 \ee
 The form of the solutions is not particularly illuminating. We show in Fig. \ref{fig:zeta4_eigenvalues} a plot of the eigenvalues $E(\nu=\{0,1,2,3\},g)$ which are the solution of the above equations, along with the numerical solution of the Schr\"odinger equation.

\subsection*{Large $\zeta$ case and an unsolved puzzle}

\begin{figure}[tbp] 
   \centering
  \rotatebox{90}{\scriptsize$\text{Re }E(g)$} \raisebox{-.41\height}{\includegraphics[width=.30\textwidth]{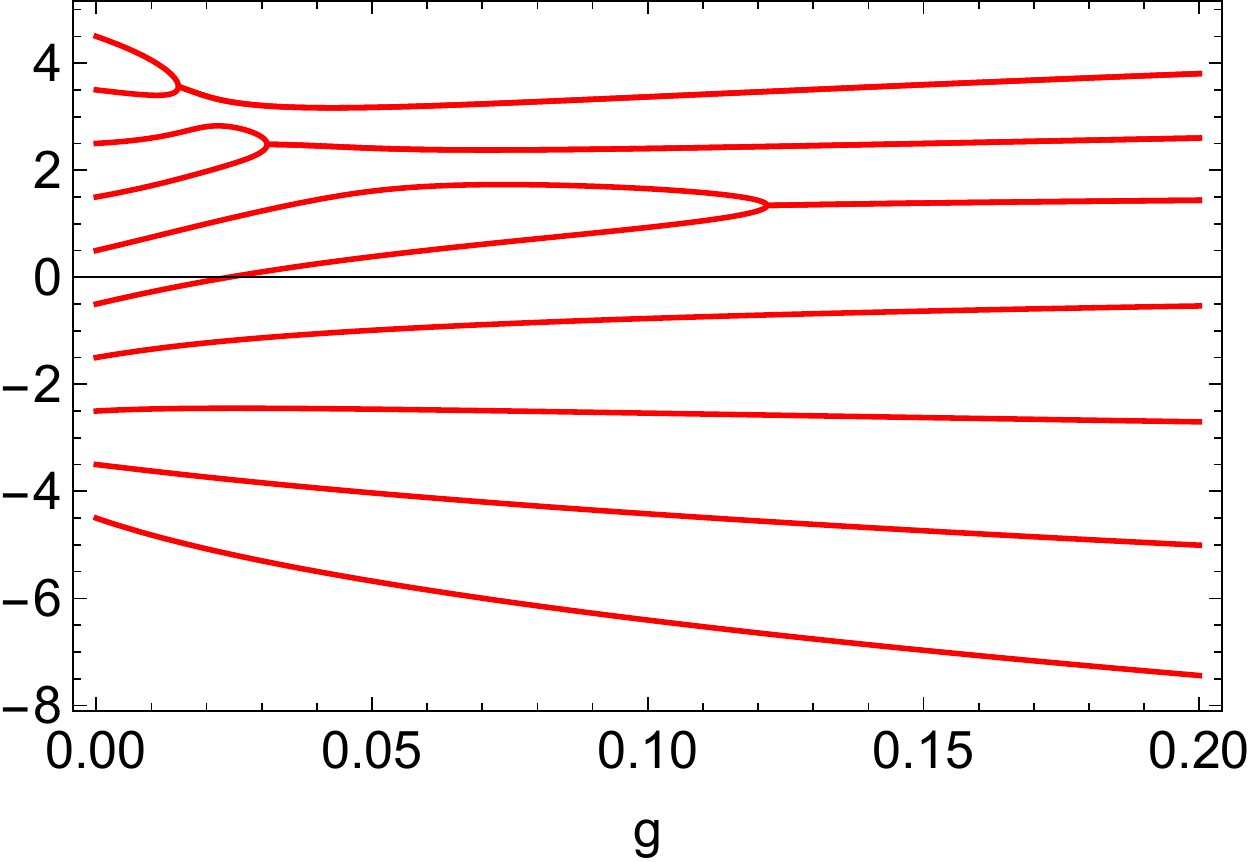}\includegraphics[width=.30\textwidth]{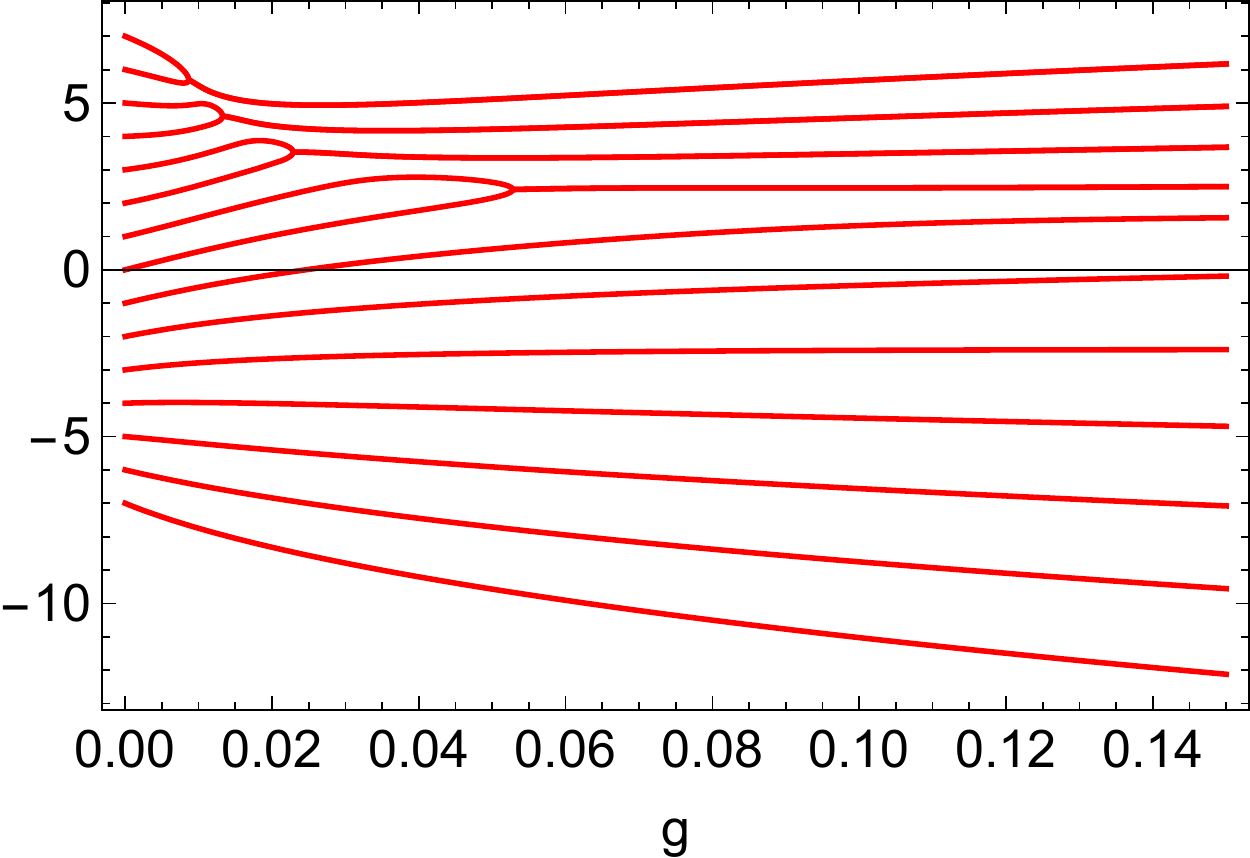}\includegraphics[width=.30\textwidth]{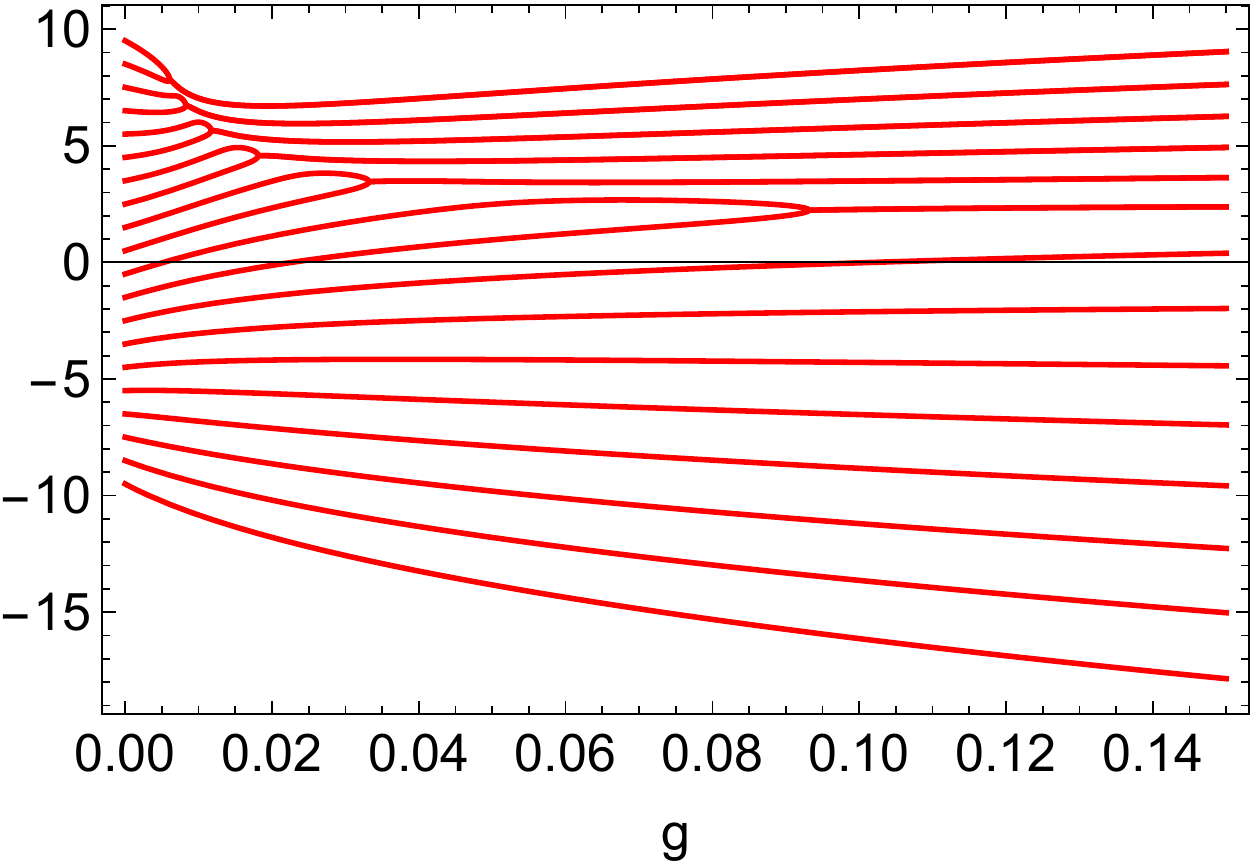}} 
   \caption{A plot of first $\zeta$ eigenvalues to all order of perturbation theory for $\zeta=10,15,20$ (left to right).}
   \label{fig:largezeta_eigenvalues}
\end{figure}

We found it amusing to also discuss briefly a case where $\zeta$ is large. Although solutions to the algebraic equation $\hat h\, u=E\,u$ do not have a nice closed form, they can nevertheless be easily plotted. In Fig. \ref{fig:largezeta_eigenvalues} we plot perturbative eigenvalues for three values of $\zeta=10,15,20$. Notice that in all cases the top lying states merge into complex-conjugate pairs at some value of the coupling $g$. It is an interesting question of how and whether these complex parts can be cured by the non-perturbative contributions. Recall that the perturbation theory in all of these cases has a perfectly finite radius of convergence. Further these imaginary parts are completely unambiguous. We suspect that non-perturbative contributions must somehow contribute a multi-valued result with an imaginary part, possibly as a result of a complete resummation of multi-instantons, in order to cancel this pathology. At this moment, however,  we do not know if this is true and present this as an open problem.

\subsection{Complex saddles and the role of the hidden topological angle}
The non-normalizable states \eqref{non-normalizable}
provides a pseudo-QES system for which all orders perturbative results are obtained. But due to non-normalizibility, these states 
are not a part of the Hilbert space, and all orders perturbative results cannot be correct expressions. 

This is puzzling from a semi-classical point of view.  Presumably, the all order perturbative result arises from the perturbative saddle, 
but  there are no real  non-perturbative saddles that can  contribute to the path integral for the lowest $\zeta$-states.  In the inverted potential,  $-V(x) =-\half (W'(x)^2+\zeta g  W''(x))$, a classical particle starting at the higher  hill-top will over shoot the lower hill top and fly of to infinity. Thus, the action of such saddles is infinite and cannot contribute to semi-classical expansion of path integral.  
As discussed in depth in \cite{Behtash:2015loa}, the resolution of this puzzle is again given by complex bions.  

The complex bion contribution to the energy for level $\nu$ is given by   (setting $A=1$ in \eqref{c-b}, note that $S_b=1/3$), one finds  
  \begin{align}
 E^{\rm n.p.}_{\pm}(\nu,g,\zeta) =  [{\cal CB}]_{\pm} =  -\frac{1}{2\pi} \frac{1}{\nu!}  \left( \frac{g}{2} \right)^{\zeta- 2 \nu -1} 
    \Gamma(\zeta - \nu)  e^{ \pm i \pi (\zeta - \nu)  } e^{-S_b/g} \left( b_0(\nu,\zeta)  + b_1(\nu,\zeta) \, g + \cdots \right),
    \label{c-b-2}
 \end{align} 
implying an  imaginary  ambiguous parts of the complex bion amplitude of the form,
 \begin{align}
{\rm Im}\, E^{\rm n.p.}_{\pm}(\nu,g,\zeta)   &=  \mp \frac{1}{2} \frac{1}{\nu!}  \left( \frac{g}{2} \right)^{\zeta- 2 \nu -1} 
   \frac{1} {\Gamma(1+ \nu -\zeta ) } e^{-S_b/g}   \left( b_0(\nu,\zeta)  + b_1(\nu,\zeta) \, g + \cdots \right).
   \label{im-cb-2}
 \end{align}
 Few comments are in order: 
 \begin{itemize}
\item For generic $\zeta$, the contribution of the complex bion is two-fold ambiguous.   This ambiguity cancels against the ambiguity in the Borel resummation of perturbation theory.   If $\zeta\in\mathbb{N}^+$, the ambiguity vanishes for first $\zeta$ states. 
 
 \item    \noindent {\bf $\bm \zeta$-odd:}   
 For level $\nu=0,2, \cdots, \zeta-1$, the complex bion increases the energy thanks to the hidden topological angle $\theta_{\mathrm{HTA}}=\pi$.  For levels $\nu=1,3, \cdots, \zeta-2 $,  the complex bion reduces  the energy.  The existence of these complex saddle gives non-perturbative contributions and  is the reason that these states are not exactly solvable.

  \item    \noindent {\bf $\bm \zeta$-even:}   For levels $\nu=0,2, \cdots, \zeta-1$, the the complex bion reduces the energy. 
 For level $\nu=1,3, \cdots, \zeta-2 $, the  complex bion increases the energy compared to all order perturbative result.   
 \end{itemize} 
Although we do not report here the details, 
all predictions arising from complex bions are realized in numerical solutions.  In particular, the deficit of all orders convergent perturbative result and the numerical solution is matched to a very high accuracy by the complex bion contribution at weak coupling.

 \subsection{Bender-Wu method for $\zeta$-deformed theory}
 The application of the Bender-Wu method to  TDW system was performed using the \texttt{BenderWu} Mathematica package of \cite{Sulejmanpasic:2016fwr}.  
Results are tabulated in Tables \ref{TDW1}, \ref{TDW2}, \ref{TDW3}   for $\zeta=1,2, 3$ for the lowest lying four levels $\nu=0,1,2,3$ in Appendix B. Further using the \texttt{BenderWu} package we are able to construct  the series as an analytic function of $\zeta$. 

 The main conclusion of these analysis are
\begin{itemize}
\item For positive integer   $\zeta$, Bender-Wu approach yields  a convergent perturbation theory for level number $\nu=0,1, \cdots, \zeta-1$. The  summation of  perturbation theory  gives precisely the same result as in the pseudo-QES approach.   These are all orders perturbative solutions to a non-perturbative problem. 
Unlike DSG,   this  is an incorrect, or rather incomplete, result.
There exists non-perturbative corrections  that arise from complex bion  saddles.  For higher states, Bender-Wu approach yields an asymptotic expansion.  
\item   For   generic $\zeta$, Bender-Wu approach yields  an asymptotic perturbation theory, which can be viewed as the leading part of the resurgent trans-series.  
\end{itemize}

\subsubsection{All orders perturbation theory }
 
 The TDW system with integer $\zeta$ shows two types of behavior in perturbation theory. For the level numbers $\nu=0, 1, \cdots, \zeta-1$ the perturbation series is convergent, and it is asymptotic 
 otherwise. Recall that we observed a similar behavior in the DSG system, where the perturbation theory summed to an exact result at an appropriate $\theta$ angle. We will see, however, that  the perturbation theory, although having a finite radius of convergence, gives an incorrect, or rather incomplete, result. In fact, we will show that the perturbation theory result is obtained exactly (i.e. to all orders) from an ansatz in the wave-function $P(x)e^{W(x)/g}$, where $P(x)$ is a polynomial of order $\zeta-1$. Such an result is clearly non-normalizable\footnote{As we discussed, it is most convenient to build the perturbation theory in the canonical normalization, in which replaces $x\rightarrow \sqrt{g}x$. The wave-function $\psi\propto e^{ W(\sqrt{g}x)/g}$ is then easily seen to be normalizable to any finite order in perturbation theory by expanding it around the global minimum $x=-a/\sqrt{g}$ . In other words the perturbation theory is oblivious to the global boundary conditions.}, and is therefore inadmissible as a solution. 

This is in contrast to the  exactly solvable states of the DSG example in which the  real bion  cancels exactly the complex bion contribution, and convergent  perturbation theory yields exact results. In the present case, there are no real saddle contribution  to cancel the complex bion contribution. The non-perturbative contribution of the complex bion is, for the ground state $\nu=0$,  at leading order of the form, 
  \begin{align}
 \Delta E^{\rm n.p.}  = - \frac{g^{\zeta-1}}{2^{\zeta}\pi}  \Gamma(\zeta)  e^{ \pm i \pi \zeta } e^{-S_b/g} +\dots\,.
 \end{align}

   If $\zeta$ is an odd integer,  $\zeta=1, 3, 5, \cdots$, the contribution of the complex bion is positive.  Note that $\zeta=1$ case is supersymmetric. For the present potential, supersymmetry is dynamically broken, and ground state energy is positive.  
   In the bosonized language of the supersymmetric theory, the positivity of the ground state energy is due to the fact that the hidden topological angle is $\theta_{\mathrm{HTA}}=\pi$. 
   
   For $\zeta=1$,  of course, perturbation theory  is convergent and gives zero to all orders for the energy for the ground state $\nu=0$.  The perturbation theory for the wave function is convergent and upon summation,  produces $\psi\propto e^{ W(\sqrt{g}x)/g}$, the non-normalizable (perturbative) solution to the Schr\"odinger equation. 
   
   For $\zeta=2$ the perturbation theory is again rather trivial for the first two levels $\nu=0,1$, giving
   \be
   E(\nu=0)= - a,  \qquad E(\nu=1)= + a, \qquad a=\half \;. 
   \ee
The reason for this is, as was pointed out in \cite{Verbaarschot:1990fa, Verbaarschot:1990ga}, that the system can be related to a two supersymmetric system with the substitution $\tilde W(x)=W(x)-\frac{1}{x\pm a}$. The ground states of these SUSY systems correspond to the ground state and the first excited state\footnote{Note that the map $\tilde W'(x)=\tilde W(x)-\frac{1}{x\pm a}$ is singular at either $x=-a$ or at $x=+a$. Because of this the map disallows decomposition of the Hamiltonian into operators } of the $\zeta=2$ tilted double well, and their perturbation theory is protected by supersymmetry.

  For $\zeta=3$,  perturbation theory  is convergent, but  does not truncate. For the lowest lying three states, few terms in the perturbative expansion and their sum gives:
 \begin{eqnarray}
 E^{\rm pert.}(\nu=0,g) & =& \textstyle{ -1-g+\frac{3 g^2}{2}-4 g^3+\frac{105 g^4}{8}-48 g^5+\frac{3003 g^6}{16}-768
   g^7+\frac{415701 g^8}{128}-14080 g^9+O\left(g^{10}\right)}   \cr
& =&  -\frac{2} {\sqrt{3}}  \cos \left(\frac{1}{3} \arccos \left(3 \sqrt{3} g\right)\right), \cr
 E^{\rm pert.}(\nu=1,g) & =& 0+      2 g+8 g^3+96 g^5+1536 g^7+28160 g^9+559104 g^{11}+O\left(g^{12}\right) \cr
& =& \frac{2}  {\sqrt{3} } \sin \left(\frac{1}{3} \arcsin\left(3 \sqrt{3} g\right)\right),  \cr
 E^{\rm pert.}(\nu=2,g) & =& \textstyle{  + 1-g-\frac{3 g^2}{2}-4 g^3-\frac{105 g^4}{8}-48 g^5-\frac{3003 g^6}{16}-768
   g^7-\frac{415701 g^8}{128}-14080 g^9- O\left(g^{10}\right)}    \cr
& =&  \frac{2}{\sqrt{3}}   \cos \left(\frac{1}{3} \arccos \left(-3 \sqrt{3} g\right)\right). 
\label{pert-result-DW}
\end{eqnarray}
The perturbation theory has a finite radius of convergence for these three lowest lying states. The  radius of convergence is 
 \begin{align}
g \leq g_c = \frac{1}{3 \sqrt 3}. 
\end{align}
Recall that $g_c$ is a branch point of 
 $ \arcsin\left(3 \sqrt{3} g\right) $ and $ \arccos\left(3 \sqrt{3} g\right) $. 

 Note that for $g<g_c$, all these three solutions are real. At $g=0$  these solutions start at $-1,  0,1$. Perturbative eigenvalue spectrum changes as a function of $g$ for  $g<g_c$,  but at $g=g_c$, two real higher eigenvalues collide and move to the complex plane, with real and imaginary parts. This perturbative conclusion is obviously incorrect, but it is not currently clear what is the mechanism which turns these complex eigenvalues of the convergent perturbation theory and QES solution  into real ones.

Unlike the text-book examples of saddles such as instantons, in which, instantons  lead to level splitting of otherwise degenerate levels,  in the present case, the complex bions lead to either up or down shift of the energy compared to all order perturbative result. 
It is still meaningful to include non-perturbative contribution, because all orders perturbative result is known exactly.

\subsubsection{Self-resurgence and the Dunne-\"Unsal relation}

Performing the Bender-Wu analysis via the \texttt{Mathematica} package \texttt{BendeWu}, we can find the sub-leading corrections to the leading factorial growth. For level $\nu$,   we have: 
\begin{align}
 a_{n}(\nu,\zeta) 
 &=   \frac{1}{2\pi}   \frac{1}{(2 )^{\zeta -2\nu-1} } \frac{1}{\Gamma(1+\nu -\zeta)}  
   \frac{(n +2\nu-\zeta )!} {(S_b)^{n-\zeta + 1}}  \cr&\qquad\times   \left[b_0(\nu,\zeta) +{ S_b\, b_1(\nu,\zeta)  \over n-\zeta+2\nu}   +  {S_b^2\,b_2(\nu,\zeta)   \over (n-\zeta+2\nu)(n-\zeta+2\nu-1)}+ \cdots\right].    
 \label{BW-P-4}
\end{align}
$b_0=1$ and where the pre-factor is constrained by demanding that the Borel sum ambiguity of the leading asymptotic growth of the perturbation series is exactly cancelled by the complex bion contribution to the ground state energy.

Further by using the \texttt{BenderWu} package~\cite{Sulejmanpasic:2016fwr} we can obtain a perturbative expansion for the energy eigenvalues as a function of coupling $g$, level number $\nu$, and parameter $\zeta$. For example, up to fourth order in $g$, we obtain an expression 

\begin{align}
 E^{\rm pert}(\nu, g, \zeta) &=a_{0}(\nu,\zeta)+a_{1}(\nu,\zeta)\,g+a_{2}(\nu,\zeta)\,g^2+a_{3}(\nu,\zeta)\,g^3+a_{4}(\nu,\zeta)\,g^4+{\cal O}(g^{5})\,,
\end{align}
where
\begin{align}\nonumber
a_{0}(\nu,\zeta)&=\nu +\frac{1}{2}-\frac{\zeta }{2},\\\nonumber
a_{1}(\nu,\zeta)&=\frac{1}{2}  \left(-\zeta ^2+6 \zeta  \nu +3 \zeta -6 \nu ^2-6 \nu -2\right),\\\nonumber
a_{2}(\nu,\zeta)&=\frac{1}{4} \Big(4 \zeta ^3-21 \zeta ^2 (2 \nu +1)+\zeta  \left(102 \nu ^2+102
 \nu +35\right)-2 \left(34 \nu ^3+51 \nu ^2+35 \nu +9\right)\Big)\\\nonumber
a_{3}(\nu,\zeta)&=\frac{1}{4} \Big(-16 \zeta ^4+123 \zeta ^3 (2 \nu +1)-2 \zeta ^2 \left(498 \nu
   ^2+498 \nu +173\right)+3 \zeta  \left(500 \nu ^3+750 \nu ^2+528 \nu
   +139\right)\Big.\\\nonumber
   &\hspace{0.4cm}\Big.-2 \left(375 \nu ^4+750 \nu ^3+792 \nu ^2+417 \nu
   +89\right)\Big),\\\nonumber
a_{4}(\nu,\zeta)&=\frac{1}{16} \Big(336 \zeta ^5-3453 \zeta ^4 (2 \nu +1)+8 \zeta ^3 \left(5010
   \nu ^2+5010 \nu +1753\right)\Big.\\\nonumber
   &\hspace{0.4cm}-6 \zeta ^2 \left(16330 \nu ^3+24495 \nu ^2+17483
   \nu +4659\right)\\\nonumber
   &\hspace{0.4cm}+\zeta  \left(106890 \nu ^4+213780 \nu ^3+230550 \nu
   ^2+123660 \nu +27073\right)\\
   &\hspace{0.4cm}-2\Big. \left(21378 \nu ^5+53445 \nu ^4+76850 \nu
   ^3+61830 \nu ^2+27073 \nu +5013\right)\Big).
\end{align}

Using the Dunne-\"Unsal relation, the knowledge of perturbative expansion around the perturbative saddle at order $g^{n}$  is sufficient to deduce the  fluctuations around the  the leading non-perturbative saddle at order $g^{n -1}$. The leading non-perturbative saddle is the complex bion.   
 The non-perturbative  
contribution to the energy for level $\nu$ is  given by 
  \begin{align}
 E^{\rm n.p.}_{\pm}(\nu,g,\zeta) =  [{\cal CB}]_{\pm} =  -\frac{1}{2\pi} \frac{1}{\nu!}  \left( \frac{g}{2} \right)^{\zeta- 2 \nu -1} 
    \Gamma(\zeta - \nu)  e^{ \pm i \pi (\zeta - \nu)  } e^{-S_b/g}   { \cal P}_{\rm fluc}(\nu, g, \zeta) 
    \label{c-b-3}
 \end{align} 
where ${ \cal P}_{\rm fluc}(\nu,g,\zeta)$ is the fluctuation operator around the complex bion  saddle.  According to the result of \cite{Dunne:2013ada,Dunne:2014bca} (see also \cite{Gahramanov:2015yxk}),   $ { \cal P}_{\rm fluc}(\nu,g, \zeta)$ is  completely dictated by $E^{\rm pert.} (\nu,g, \zeta)$ in a constructive way. 
    \begin{align}
{ \cal P}_{\rm fluc}(\nu, g, \zeta) & =  \frac{\partial{E^{\rm pert.}}}{\partial \nu}  \exp\left[ S_b  \int_0^g  \frac{dg}{g^2} \left(  \frac{\partial{E^{\rm pert.}}}{\partial \nu}  -  1 +  \frac{2g (\nu+ \frac{1}{2} - \frac{\zeta}{2}) }{S_b}   \right) \right]    \cr
& = b_0(\nu,\zeta) +   b_1(\nu,\zeta) \,  g  + b_2(\nu,\zeta)\,    g^2  +  b_3(\nu,\zeta) \,  g^3  +\dots,
\label{fluc-5}
\end{align}
where 
\begin{align}\label{eq:TDW_bcoeff_DU}
 b_0(\nu, \zeta) &=1, \cr  
 b_1(\nu, \zeta) &=\frac{1}{6} \left(-21 \zeta ^2+3 \zeta  (34 \nu +23)-6 \nu  (17 \nu +23)-53\right), \cr
  b_2(\nu, \zeta) &=   \frac{1}{72} \Big( 441 \zeta ^4-36 \zeta ^3 (119 \nu +60)+3 \zeta ^2 \left(4896 \nu ^2+4632 \nu +973\right)\cr&\hspace{0.4cm}-6 \zeta  \left(3468 \nu ^3+4788
   \nu ^2+1898 \nu +13\right)+10404 \nu ^4+19152 \nu ^3\cr&\hspace{0.4cm}+11388 \nu ^2+156 \nu -1277
 \Big), \cr
   b_3(\nu, \zeta)&=\frac{1}{1296}\Big(-9261 \zeta ^6+567 \zeta ^5 (238 \nu +79)-162 \zeta ^4 \left(4879 \nu ^2+2883 \nu +143\right)\Big.\cr&\hspace{0.4cm}+27 \zeta ^3 \left(87856 \nu
   ^3+70704 \nu ^2+4014 \nu -4929\right)\cr&\hspace{0.4cm}-18 \zeta ^2 \left(213282 \nu ^4+214524 \nu ^3+7605 \nu ^2-45093 \nu -1465\right)\cr&\hspace{0.4cm}+9
   \zeta  \left(353736 \nu ^5+431460 \nu ^4+6336 \nu ^3-181836 \nu ^2+5794 \nu +45941\right)\cr&\hspace{0.4cm}-1061208 \nu ^6-1553256 \nu
   ^5-28512 \nu ^4+1091016 \nu ^3\cr&\hspace{0.4cm}-52146 \nu ^2-826938 \nu -336437 \Big).
\end{align}

Setting level  number $\nu=0,1$, one obtains for the fluctuations around the complex bion event contribution to the ground state 
and first excited state energies as 
\begin{align}\label{eq:TDW_bcoeff_DU-2}
 b_0^{\nu=0}(\zeta) &=1 \\
 b_1^{\nu=0}(\zeta) &= \frac{1}{6}  \left(      -53 +   69  \zeta -  21 \zeta^2   \right)  \cr
  b_2^{\nu=0}(\zeta) &=   \frac{1}{72} \Big(-1277  - 78  \zeta + 2919   \zeta^2  -2160 \zeta^3 +   441 \zeta^4
\Big),  \cr
  b_3^{\nu=0} (\zeta) &=  \frac{1}{1296} \left(-9261 \zeta ^6+44793 \zeta ^5-23166 \zeta ^4-133083 \zeta ^3+26370 \zeta
   ^2+413469 \zeta -336437\right)   \nonumber
\end{align} 
and
\begin{align}\label{eq:TDW_bcoeff_DU-3}
 b_0^{\nu=1}(\zeta) &=1 \\
 b_1^{\nu=1}(\zeta) &= \frac{1}{6} \left(-21 \zeta ^2+171 \zeta -293\right)  \cr
  b_2^{\nu=1}(\zeta) &=   \frac{1}{72} \left(441 \zeta ^4-6444 \zeta ^3+31503 \zeta ^2-61002 \zeta +39823\right),  \cr
  b_3^{\nu=1} (\zeta) &=  \frac{1}{1296} \left(-9261 \zeta ^6+179739 \zeta ^5-1280610 \zeta ^4+4256415 \zeta ^3-6999354 \zeta ^2+5952879 \zeta -2767481\right)   \nonumber
\end{align}

On the other hand, we can find these coefficients $b_1,b_2,b_3,\cdots$ approximately from the explicit calculation of the perturbation theory via the procedure described in Section \ref{sec:cor}. We get for $\nu=0$
\begin{align}\label{eq:TDW_bcoeff_BW_nu0}
b_1(\nu=0,\zeta)=&-8.833333332924 + 11.49999999835448 \zeta - 
 3.499999997041193 \zeta^2 +o(10^{-9} \zeta^3),  \cr
 b_2(\nu=0,\zeta)=&-17.736112193 - 1.0833289712 \zeta + 40.5416588043985 \zeta^2 \cr&\hspace{0.0cm} - 
 29.99999157670543 \zeta^3+ 6.12499401506749 \zeta^4 + 
 o(10^{-6} \zeta^5),\cr
 b_3(\nu=0,\zeta)=&-259.595 + 319.03 \zeta + 20.3565 \zeta^2 - 102.697 \zeta^3 - 
 17.8679 \zeta^4 + 34.559 \zeta^5 \cr&\hspace{0.0cm}- 7.14458 \zeta^6 + o(10^{-4}\zeta^7),
\end{align}
and for $\nu=1$
\begin{align}\label{eq:TDW_bcoeff_BW_nu1}
b_1(\nu=1,\zeta)=&-48.833333096294 + 28.49999939463613 \zeta - 
 3.499999304322683 \zeta^2 +o(10^{-7} \zeta^3),  \cr
 b_2(\nu=1,\zeta)=&553.096586547 - 847.2483729877 \zeta + 
 437.5397921020106 \zeta^2 - 89.4987081484367 \zeta^3 \cr&+ 
 6.12440133941877 \zeta^4+o(10^{-4}\zeta^5),\cr
 b_3(\nu=1,\zeta)=&-2134.65 + 4591.34 \zeta - 5398.5 \zeta^2 + 3282.73 \zeta^3 - 
 987.407646031719 \zeta^4 \cr&+ 138.4467901243447 \zeta^5 - 
 7.08375962476414 \zeta^6+ o(10^{-2} \zeta^7).
\end{align}
The reader is welcome to check that \eqref{eq:TDW_bcoeff_BW_nu0} and \eqref{eq:TDW_bcoeff_BW_nu1} are numerically consistent with \eqref{eq:TDW_bcoeff_DU-2} at $\nu=0$ and  with  \eqref{eq:TDW_bcoeff_DU-3}  at $\nu=1$ within the relative error 0.05\% and 0.8\%, respectively. 

\section{Connection to Quantum Field Theory}\label{sec:QFT}
Before conclusions we take an opportunity to comment on the potential significance of these systems and one of our motivation in studying them.

On the one hand, these quantum mechanical systems are helping us establish the rules of an all orders semi-classical expansion (i.e. exact semi-classics). On the other hand, these systems have remarkable similarities with some quantum field theories, in particular, to gauge theories \cite{Unsal:2007jx} and non-linear sigma models\footnote{For an explanation of this connection and proof of volume independence in the $CP^{N-1}$ model see \cite{Sulejmanpasic:2016llc}.} \cite{Dunne:2012zk} with matter fields.

In the present context, $\zeta \in (-n_f, -n_f+1, \ldots, n_f) $-deformed theories arise as  the sectors of a multi-flavor theory 
with one bosonic position field $x(t)$ and multiple  Grassmann valued-fields $\psi^I(t)$ where $I=1,\ldots  n_f$.  For $n_f= 1$, this theory is supersymmetric for appropriate choice of couplings.  We can do a similar construction in gauge theories and sigma models. First,  we can promote a bosonic theory into  a supersymmetric one
by adding a Grassmann valued quantum field with the right gauge, global,  Lorentz quantum numbers and interactions. Then, we can replicate the fermionic sector and obtain the multi-flavor version. 
 This procedure is sketched in Quantum Mechanics as well as in the Yang-Mills and $CP^{N-1}$ theories below
\begin{align*}
\begin{array}{rlrrrrr}
&\text{Bosonic}&&\text{SUSY}&&\text{multi-flavor}&\text{generalization}\\\hline\\
{\bf QM:}   \qquad &  x(t)   & \longrightarrow  & (x(t),  \psi(t))   &\longrightarrow & (x(t),  \psi^I(t))\;,& I=1,\dots, n_f  \\\
{\bf Yang \; Mills:}   \qquad &   A_{\mu} & \longrightarrow &(A_{\mu}, \psi_{\alpha})  & \longrightarrow  &(A_{\mu}, \psi_{\alpha}^I)\;, &I=1,\dots, n_f \\
{\bf {{CP}}^{N-1}:} \qquad  & z_i  &\longrightarrow & (z_i, \psi_i)  & \longrightarrow  &(z_i, \psi^I_i)\;,& I=1,\dots,n_f
\end{array}
\end{align*}
There is currently building up evidence that this class of ``replica theories" has some number of very special properties, similar to supersymmetric ones. 

\begin{itemize} 
\item {
Consider the twisted (or graded) partition function 
\be
\tilde Z(L)=\tr \,e^{-HL}(-1)^F,
\ee
where $F$ is a fermion number. In supersymmetric theories, this is the supersymmetric Witten index \cite{Witten:1982df}. It is  an invariant  quantity independent of $L$. 
In our   multi-flavor QM system with odd number of $n_f$,  $\tilde Z(L)=0$ either for the supersymmetric theories with $n_f=1$, 
as well as non-supersymmetric theories with $n_f=3, 5, 7, \ldots$.\footnote{Vanishing index for the supersymmetric case either imply 
absence of supersymmetric ground states or Bose-Fermi paired supersymmetric ground states. Our TDW is an example of the former and DSG is an example of the latter.} This vanishing of course implies an exact spectral cancellation over the whole 
spectrum.\footnote{For DW, $I_W=0=0-0$ because there is neither bosonic nor fermionic supersymmetric ground states,  and in the SG, $I_W=0=1-1 $ because there is a Bose-Fermi paired set of ground states. In both cases, the non-zero spectrum exhibits spectral cancellation.}

In similarly constructed QFT, in particular, in  QCD(adj), at large-$N$ limit, $\tilde Z(L)$ satisfies volume independence, namely 
 \begin{align} 
\frac{ \partial{\tilde Z(L)}}{\partial L} \Big|_{N=\infty}= 0 
\end{align}
 In particular, there are no phase transitions as a function of $L$. In supersymmetric  theory, it is known that this is due to exact spectral cancellations (modulo ground states for ${\cal N}=1$ SYM, which gives the index $I_W= \tilde Z(L) =N$), a consequence of supersymmetry.  In the non-supersymmetric theories, especially, the absence  of the confinement/deconfinement phase transition \cite{Unsal:2007jx} is extremely intriguing, and points to spectral cancellation even in non-supersymmetric theory  with a  potentially emergent fermionic symmetry at large-$N$.\footnote{The existence of an emergent  fermionic symmetry  is not  ruled out by Coleman-Mandula theorem, because $N=\infty$ is free in terms of 
 hadrons, and has trivial S-matrix.}
  To see this, note that 
if the factor $(-1)^F$ is dropped, the system has a thermal interpretation and undergoes a confinement/deconfinement transition at some $L=L_c$ due to the Hagedorn growth of the density of states of large-$N$ theory.   The lack of Hagedorn instability implies that there must exist an 
extreme spectral cancellation between bosonic and fermionic sectors,  pointing to an emergent fermionic symmetry at large-$N$ limit \cite{Basar:2013sza,Basar:2014jua,Basar:2015asd}.  In this sense, our multi-flavor QM systems may form prototype for much complicated QFTs, such as QCD(adj) and two dimensional sigma models with multi-fermions.}

\item{ In multi-flavor theories, there also exist real and complex bions, correlated instanton-anti-instanton pairs. In QCD(adj), the hidden topological angle associated with complex bion is $\theta_{\rm HTA}= (4n_f-3)\pi $,  \cite{Behtash:2015kna},   In particular,  $\theta_{\rm HTA}$ is quantized for integer values of the number of flavors. This implies, as shown in   \cite{Behtash:2015kna}, 
 that the non-perturbative contributions to the gluon condensate (and by trace anomaly to the vacuum energy)  that arise from neutral and magnetic bions interfere with each other  and their total contribution vanishes.  This is identical mechanism with the \eqref{rb-cb}, where cancellation between the real and complex bion takes place. This also suggest that perturbation theory for QCD(adj) may have finite radius of convergence for some special set of states. Further analogous multi-flavor $CP^{N-1}$  systems show similar behavior \cite{Dunne:2012zk}.}
 
 \item{ More concretely, 
 the DSG quantum mechanics is related to certain twisted compactification 
of two dimensional sigma models, and is connected to them via adiabatic continuity. In particular, it 
appears as the low energy limit of two dimensional $SU(2)$ principle chiral model  and $O(3)$ model with  fermions on a small circle limit \cite{Cherman:2014ofa,Dunne:2012zk}. For $n_f=1$ flavor susy theory, the cancellation between the real and complex bion correspond to the vanishing of the spin wave condensate in the field theory.  This quantum mechanics also corresponds to   low energy limit of circle compatified $\eta$-deformed principle chiral model for a special choice of parameters \cite{Demulder:2016mja}.}
\end{itemize}

We interpret the existence of both real and complex non-BPS saddles, the quantization of hidden topological angles, and the exact spectral cancellation as useful analogs between the quantum mechanical  systems we study and the QFTs with the structure given above. While the situation in quantum field theories in general is undoubtedly much more subtle, it is useful 
to keep the   remarkable similarities between these two cases, and investigate it further.

\section{Conclusion and Outlook}\label{sec:conclusions}
This work is a step towards exact semi-classical treatment of path integral, and reveals surprising resurgent relations 
between the  perturbation theory  around the perturbative vacuum and around non-perturbative complex saddles. We were able to treat  a class of theories parametrized by $\zeta$ in a unified manner,  where $\zeta=0$ is the bosonic, $\zeta=1$ is supersymmetric, $\zeta=2, 3, \ldots$ are either QES or pseudo-QES, and generic values of $\zeta \in \mathbb R$  are also equally interesting. 
Using the Bender-Wu method, we computed the perturbative coefficient of these theories as a function of $\zeta$ and the level number $\nu$. By computing the perturbative coefficients explicitly we checked that the large-order asymptotic growth of the perturbation theory is correctly described by the early terms of the perturbative fluctuations around the complex-bion saddle via traditional resurgence, a late-term early-term correspondence. 

For both systems we study whenever $\zeta\in\mathbb{N}^+$ the factorial growth of the perturbation theory vanishes for the first $\zeta$ states. 
Using the technique of QES, we analytically show that this perturbation theory for the first $\zeta$ states converges. This all-order perturbative solution gives an exact solution if it satisfies the correct boundary condition, but otherwise suffers from non-perturbative correction. 
There was a long-standing puzzle in the QES literature about this subtlety: the perturbative solution gives an exact solution while there exists a real non-perturbative classical solution, called a real bion, in one case, and the non-perturbative correction exists while a real bion is absent in the another case. 

By analyticity in $\zeta$ we conclude that the effect of complex saddle, called complex bion, is present for $\zeta\in\mathbb{N}^+$ without any imaginary ambiguities, a phenomenon which we call the Cheshire Cat resurgence. 
We find that contributions of real and complex bions must be canceled in order for the convergent perturbative solution giving an exact answer. 
This emphasizes the importance of complex bion in the semiclassical analysis. 

We also consider about the unconventional type of resurgent relation --the self-resurgence. In the double sine-Gordon and titled double-well cases, early terms of the expansion around the perturbative saddle give sufficient information about early terms of the expansion around complex bions. By exploiting the traditional resurgence, this means that early terms of the perturbative series know about late terms of the same series: i.e. the perturbative expansion is self-resurgent.  We checked the self-resurgent property by explicitly computing the perturbative series, and found an astounding agreement. 

It is an important future study to understand the effect of complex bions in the semiclassical analysis from the viewpoint of the path integral expression. Application of the Picard--Lefschetz theory to the (UV and IR regularized) path integral gathers much attention for numerical study of lattice field theories in order to tame the sign problem~\cite{Cristoforetti:2012su, Cristoforetti:2013wha, Fujii:2013sra, Aarts:2014nxa}. 
If the classical action takes complex values, then there exist situations where interference of multiple complex classical solutions are important for physical observables~\cite{Witten:2010cx, Dumlu:2010ua, Dumlu:2011rr, Tanizaki:2014xba, Cherman:2014sba, Tanizaki:2015rda, Fujii:2015bua, Alexandru:2015xva, Hayata:2015lzj, Alexandru:2015sua, Alexandru:2016gsd}, which may remind us interference between real and complex bions. 
However, the models in this study do not have the sign problem since the classical action is a real functional at least when the coupling is physical\footnote{A sign problem does exist in the formulation of \eqref{lag-2}, as the ``Dirac operator'' determinant is not positive definite. The study of this systems however can be reduced to the study of the $\zeta$-deformed systems with $\zeta=-n_f/2,\dots, n_f/2$, all of which do not posses the sign problem.}. 
This poses an interesting question on how we can understand the contribution of complex bions with nonzero HTA based on the Lefschetz-thimble decomposition of path integral. Quite possibly, some conditions on the standard Lefschetz-thimble approached must be relaxed to accommodate the complex bion contribution  to path integral. 

Finally,  as we have already explained in Section \ref{sec:QFT},  quantum mechanics studied in this paper has a formal similarity with  multi-flavor QCD with adjoint fermions,  and some nonlinear sigma models with fermions. These theories  possess some interesting properties as supersymmetric field theories, much like our $\zeta=3, 5, \ldots$ theories possessing most of the properties of the $\zeta=1$ supersymmetric theory. 
By comparing properties of the hidden topological angle, we can speculate that magnetic and neutral bions in  QCD(adj) correspond to real and complex bions in quantum mechanics. 
It is an interesting topic to understand the non-perturbative dynamics of  gauge theories by discussing constructive or destructive interference of real and complex bions~\cite{Dunne:2016nmc}. 
It is also a great task to explain the relation between generic asymptotic nature of the perturbation theory,  quantization of the 
hidden topological angle, and potentially, convergent perturbation theory for a subset of states  in quantum field theories~\cite{Hooft:1977am, Cherman:2014ofa, Cherman:2013yfa}. 

\begin{acknowledgments}
We are grateful to Gerald Dunne, Edward Shuryak and  Sasha Turbiner for their input and comments. TS is specially thankful for the discussions with Michael Berry which elucidated Dingle's self-resurgence formula. The work of M.\"U. was supported by the DOE grant DE-SC0013036. The work of T.S. was partly supported by the DOE grant DE-FG02-03ER41260. 
The work of C.K. is supported by Center for Mathematical Sciences and Applications at Harvard University. Y.T. was supported by Grants-in-Aid for JSPS fellows (No.25-6615) when this work started and is now supported by Special Postdoctoral Researchers Program of RIKEN.  
The work of Y.T. was also partially supported by  the RIKEN iTHES project, and by the Program for Leading Graduate Schools, MEXT, Japan. M.\"U.'s  work was partially supported by the Center for Mathematical Sciences and Applications (CMSA) at Harvard University, where this work has begun.

\end{acknowledgments}

\appendix

\newpage
\section{Tables for perturbative coefficients}
We here show tables of perturbative coefficients of the DSG and TDW systems at $\zeta=1$, $2$, and $3$. 
\begin{table}[!h]\centering
\subfloat[DSG $\zeta=1$]{
\input{DSG_zeta1.tab}
\label{DSG1}}\vspace{0.cm}
\subfloat[DSG $\zeta=2$]{
\input{DSG_zeta2.tab}
\label{DSG2}}  
\subfloat[DSG $\zeta=3$]{
\input{DSG_zeta3.tab}
\label{DSG3}}
\caption{Tables of perturbative coefficients of the DSG system at $\zeta=1$, $2$, and $3$.}
\end{table}
\begin{table}[!h]\centering
\subfloat[TDW $\zeta=1$]{
\input{TDW_zeta1.tab}
\label{TDW1}}\vspace{0cm}
\subfloat[TDW $\zeta=2$]{
\input{TDW_zeta2.tab}
\label{TDW2}}
\subfloat[TDW $\zeta=3$]{
\input{TDW_zeta3.tab}
\label{TDW3}}
\caption{Tables of perturbative coefficients of the TDW system at $\zeta=1$, $2$, and $3$.}
\end{table}

\clearpage
\bibliographystyle{JHEP}
\bibliography{./bibliography}
\end{document}